\documentclass[pra,twocolumn,showpacs,superscriptaddress,floatfix,footinbib,amsmath,amssymb]{revtex4}

\usepackage[utf8]{inputenc}
\usepackage{graphicx}% Include figure files
\usepackage[normalem]{ulem}
\usepackage[usenames,dvipsnames]{color}
\usepackage[nointegrals]{wasysym}
\usepackage{commands}
\usepackage{cleveref}
\usepackage{upgreek}
\usepackage{amsmath}

\newcommand{\id}{\mathbb{1}}

\begin{document}

\title{Coherent Electron Optics with Ballistically Coupled Quantum Point Contacts}

\author{J. Freudenfeld}\affiliationPDI
\author{M. Geier}\affiliationFUtheory
\author{V. Umansky}\affiliationWIS
\author{P. W. Brouwer}\affiliationFUtheory
\author{S. Ludwig}\affiliationPDI

\date{\today}

\pacs{
}
\begin{abstract}
The realization of integrated quantum circuits requires precise on-chip control of charge carriers. Aiming at the coherent coupling of distant nanostructures at zero magnetic field, here we study the ballistic electron transport through two quantum point contacts (QPCs) in series in a three terminal configuration. We enhance the coupling between the QPCs by electrostatic focusing using a field effect lens. To study the emission and collection properties of QPCs in detail we combine the electrostatic focusing with magnetic deflection. Comparing our measurements with quantum mechanical and classical calculations we discuss generic features of the quantum circuit and demonstrate how the coherent and ballistic dynamics depend on the details of the QPC confinement potentials.
\end{abstract}

\maketitle

Quantum point contacts (QPCs) are the smallest fundamental units of solid state based quantum circuits. These short tunable one-dimensional (1D) constrictions in a two-dimensional electron system (2DES) display an astonishingly rich spectrum of physics from the famous conductance quantization \cite{Landauer1981,QPCsBeenakker1988,QPCsWharam1988} to many-body interaction effects such as the so-called 0.7-anomaly \cite{Lunde2009,Micolich2011,Bauer2013}. Individual QPCs are important components in quantum circuits, e.g., as charge detectors \cite{Field1993} or to split quantum-Hall edge channels \cite{Yang2003,Prokudina2014}. The complexity of QPCs has been revealed in many experiments \cite{Micolich2011} including shot noise measurements \cite{Hashisaka2008}, scanning gate spectroscopy \cite{Topinka2000,Brun2014}, thermoelectric studies \cite{vHouten1992}, phototransport \cite{Roessler2008a}, magnetotransport out of equilibrium \cite{Williamson1990,Chen2013}, or quantum transport through freely suspended devices \cite{Roessler2010}. Aspects of the ballistic dynamics of coupled QPCs have been studied in experiments focusing on non-ohmic resistance \cite{Beton1989,Wharam2QPCs1988,Kouwenhoven1989,Coleridge1994,Liu2010} or magnetic deflection \cite{Houten1989,Williamson1990,Chen2013}, spin-orbit coupling \cite{Chesi2011,Shun-Tsung2017}, defect scattering \cite{Koonen2000} or diffraction at a QPC \cite{Khatua2014}. In ballistic quantum circuits, QPCs could serve as a coherent electron source or sink. However, such an utilization requires a comprehensive understanding of the QPCs carrier emission and collection properties. Both are characterized by the coupling between the QPC's local 1D modes and the ballistic dynamics in the 2DES. Here we study the combined ballistic and coherent dynamics of two QPCs in series. Our results substantially improve our understanding of QPCs and provide a viable basis for the design of ballistic quantum circuits.

We consider two QPCs, defined electrostatically using the usual split gate design. They are tuned to their quantized conductance regimes and interact via the exchange of ballistic electrons via a free, i.e., grounded, region of 2DES. We demonstrate that the mutual coupling can be strongly enhanced by fine tuning an electrostatic lens \cite{Sivan1990,Spector1990} between the two QPCs. The lens functions by refocusing carriers diverging from one QPC into the second QPC. For studying this \textit{electrostatic focusing} we combine it with \textit{magnetic deflection} \cite{Tsoi1974,Houten1989} in a field perpendicular to the 2DES. (We avoid the common term \textit{magnetic focusing}, as a homogeneous magnetic field merely deflects currents.) This combination is essential to fully determine the angular resolved emission spectrum of the QPCs and explore electrostatic focusing between QPCs.
Our magnetic fields are so small that we can neglect the Zeeman splitting of electron states.

\begin{figure}[ht]
\includegraphics[width=1\columnwidth]{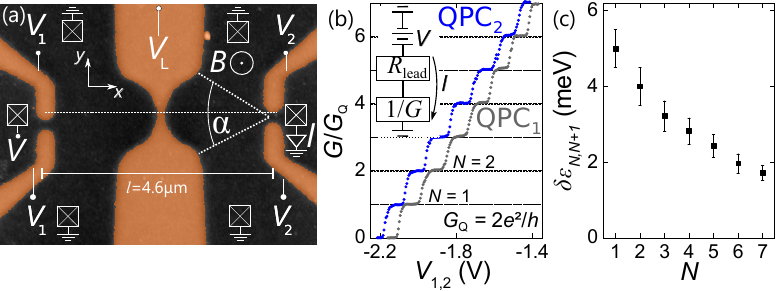}
\caption{(a) Atomic force microscope image of the sample; Ti/Au gates on GaAs surface (dark). Gate voltages $V_1, V_2, V_\text{L}$ are used to define in the 2DES below QPC$_{1,2}$ and a lens. Source-drain voltage $V$ is applied across QPC$_1$; current $I$ is measured through QPC$_2$; the region in between is grounded via 4 ohmic contacts (squares). The horizontal dashed line is the principal axis of the lens with aperture angle $\alpha\simeq55^\circ$.
(b) Individual linear response pinch-off curves $G(V_{1,2})$ of QPC$_{1,2}$ ($V_{2,1}=V_\text{L}=0$), corrected for lead resistance $R_\text{lead}$.
(c) Energy spacings between subsequent subbands.
}
\label{fig:sample}
\end{figure}
For our model calculations we first define a 2D electrostatic potential landscape based on the actual sample layout and characterization measurements. Then we determine the ballistic electron dynamics by numerically solving either the Schr\"odinger equation or the classical equation of motion. 
For our measurements we use an (Al,Ga)As/GaAs heterostructure containing a 2DES 107\,nm beneath its surface. Figure \ref{fig:sample}(a)
displays the surface including metal gates used to define the two QPCs and a lens in between. The 2DESs Fermi energy and mean free path measured at cryogenic temperatures are $\ef^0\simeq10.9$\,meV and $l_\text m\simeq24\,\mu$m. We performed direct current (dc) measurements in a helium-3 evaporation cryostat at $T\simeq250$\,mK. For a basic characterization we present in \fig{fig:sample}{b} linear response pinch-off curves of the individual QPCs. The conductance as a function of gate voltages $V_1, V_2$ features flat plateaus at $N G_\text Q$ with $N=1,2,\dots$ and the spin degenerate 1D conductance quantum $G_\text Q =2e^2/h$.

The precise relation between the confinement potential of the QPCs and carrier emission profile is central for understanding the ballistic carrier dynamics and for optimizing a quantum electronic circuit. The lateral confinement defines the mode structure of the 1D channel while its potential shape in current direction [$x$-axis in \fig{fig:sample}a] governs the coupling of the 1D modes into the surrounding 2DES. 
Our pinch-off curves exhibit smooth steps between conductance plateaus suggesting reflectionless transmission between the free 2DES and the QPCs. This indicates smooth (parabolic) potential barriers as also implemented in our model \cite{Heyder2015}. Importantly, for reflectionless coupling the lateral 1D eigenmode structure is preserved in the coherent QPCs' emission profile.

Although the conductance steps of our QPCs in \fig{fig:sample}b are almost equidistant as a function of gate voltages, the corresponding energy spacings between the 1D subbands [cf. \fig{fig:sample}c] strongly decrease with $N$. These energies are incompatible with parabolic lateral confinement for $N\ge4$ \cite{Geier2020-1}. They point to a transition by screening from a parabolic confinement for $N\le1$ towards a hard wall potential for $N\ge4$ \cite{Laux1988}.

In the following measurements we apply a dc voltage of $V=-1\,$mV across one QPC (emitter) and measure the current $I$ flowing to ground through the second QPC (detector), cf.\ \fig{fig:sample}{a}. Electrons move ballistically between the QPCs as their distance of $l\simeq4.6\,\mu$m is smaller than $l_\text m$. Alternative current paths include backscattering through the emitter or scattering to the grounded side contacts ($\iside$), such that the emitter current $I_\text{em}=\iside+I$. The resistance between the center region and ground at the side contacts is $\simeq$\,$37\,\Omega$, small compared to the QPC resistances exceeding $1.8$\,k$\Omega$ in our measurements for $N\le7$. Nevertheless, backscattering from the macroscopic side contacts causes a small shift of the local chemical potential between the QPCs and, hence, a small diffusive contribution to the detector current $I$, such that $I=\ibal+\idif$ with $\idif<0.02 \, \iside$ (\idif\ is additionally influenced by a tiny voltage offset of the current amplifier), cf.\ Ref.\ \cite{supplement}. Here, we are interested in \ibal, the contribution to the detector current generated by carriers moving ballistically between emitter and detector.

\begin{figure*}
\includegraphics[width=2.0\columnwidth]{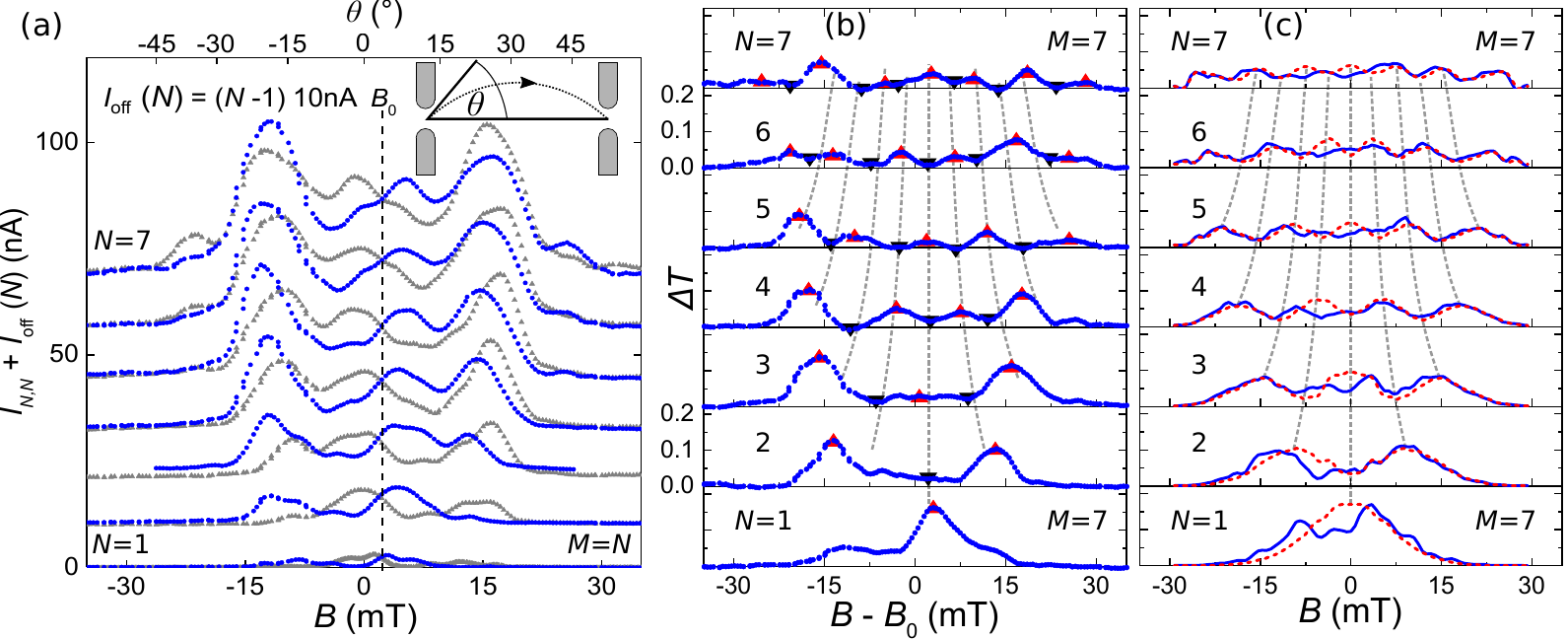}
\caption{
(a) Magnetic deflection with two QPCs in series: measured detector current $I_{N,N}$ versus perpendicular magnetic field $B$ for both QPCs tuned to the $N$th conductance plateau with $N=1,...7$. Two data sets (gray, blue) correspond to opposite current directions; vertical shifts $I_\text{off}(N)$ for clarity. %The inset is a sketch of the experiment. 
Measured in (b) versus calculated in (c) transmission differences $\Delta T_{N,M=7}$. Model curves in (c) for perfect symmetry and zero lens potential (red dashed) and with corrections of the QPC positions and accounting for the piezoelectric dip of the lens potential (solid blue lines).
Maxima and minima of $\Delta T_{N,M=7}$ are marked in panel (b) with red (black) triangles. Dashed gray lines [identical in (b) and (c)] connect the $n$th maxima for odd (even) $N$ and the $n$th minima for even (odd) $N$.} 
\label{fig:magnetic_focusing}
\end{figure*}
\ibal\ is limited by the divergence of the carrier modes emerging from a QPC: carriers are emitted within an aperture angle which depends on the height of the barrier in current direction and the lateral confinement along it. Given their divergence most of the carriers miss the detector and  mostly contribute to $\iside$. The purpose of our lens is to re-focus these carriers to enhance the coupling between the QPCs. 

To first characterize the divergence and lateral mode structure of the QPCs we perform magnetic deflection experiments without electrostatic focusing \cite{Koonen2000,Khatua2014}. Our QPCs are aligned in series, such that ballistic carriers emitted at a larger angle reach the detector at a higher field. In \fig{fig:magnetic_focusing}a
we display example curves $I_{N,N}$ with both QPCs tuned to the center of the $N$th conductance plateau with $N=1,\dots,7$. The two sets of curves correspond to opposite current directions, i.e.\ exchanged role of emitter vs.\ detector. The symmetry is predicted by the Onsager-relations \cite{Casimir1945} for a multi-terminal device, here $\left.I(B)\right|_\leftarrow=\left.I(-B)\right|_\rightarrow$, where arrows indicate the opposite current directions \footnote{An offset of the symmetry point (vertical dashed line) is caused by the perpendicular component of a small residual field of $B_0\simeq2.57\,$mT originating from magnetized connector pins nearby the sample. Deviations from $\left.I(B)\right|_\leftarrow=\left.I(-B)\right|_\rightarrow$ can be attributed to the in-plane component of the residual field or magnetic impurities.}. Below, we will decipher the information the serial current $I(B)$ encodes on the mode-to-mode coupling between the QPCs.

To predict \ibal\ we numerically solve the Schr\"odinger equation for a single electron moving in a 2D Fermi gas connected to leads as indicated in \fig{fig:sample}a. We account for the estimated dephasing length of $l_\phi\simeq0.5\,\mu$m by energy averaging, where $l_\phi$ is dominated by temperature and bias broadening. To mimic the measured mean free path of $l_\text m\simeq24\,\mu$m we include a weak homogeneous absorbing potential between the QPCs. To model the lateral confinement of the QPCs we use a hard wall potential, where its gate voltage dependent width and depth (at the center of the constrictions) are determined from the subband spacings plotted in \fig{fig:sample}c  \cite{Geier2020-1}. The opening of the constrictions towards the leads follow the semicircular shape of the gates. The lens potential controlled by the gate voltage $\vl$, cf.\ \fig{fig:sample}{a}, is added on demand. From the solution of the Schr\"odinger equation we extract the total transmission probability $T_{N,M}(B,\vl) = \sum_{n,m} t_{n,m}$ of a ballistic and coherent electron through the two QPCs in series with the first (second) QPC set to the $N$-th ($M$-th) conductance plateau. Thereby $t_{n,m}(B,V_{\rm L};V_1,V_2)$, $n=1,2,\ldots,N$, $m=1,2,\ldots,M$ are the transmission probabilities between the occupied transverse eigenmodes of the detector and emitter. We consider slowly varying QPC potentials and neglect coherent backreflections into the QPCs. In this limit $t_{n,m}$ do not depend on the gate voltages for $n \le N$ and $m \le M$ and can be reconstructed from the total transmissions as $t_{n,m} = T_{N,M} - T_{N-1,M}- T_{N,M-1} + T_{N-1,M-1}$. The Landauer formula relates $T_{N,M}(B,\vl)$ to the measured ballistic current, $I_{N,M}^\text{ball}(B,\vl) = G_\text Q V\,T_{N,M}(B,\vl)$.

For better illustrating the mode structure we have also measured $I_{N,M=7}(B)$ for $1\le N\le7$ with fixed $M$. Aiming at a direct comparison with model predictions we subtract the $B$-field independent \idif\ from the raw data obtaining $I_{N,M=7}^\text{ball}(B)=I_{N,M}(B)-I_{N,M}^\text{diff}(B)$ \cite{supplement}. In \fig{fig:magnetic_focusing}b we plot the measured transmission differences, $\Delta T_{N,M=7}(B)\equiv\left[I_{N,M=7}^\text{ball}(B)-I_{N-1,M=7}^\text{ball}(B)\right]/G_\text QV$, and in \fig{fig:magnetic_focusing}c as red dashed lines the bare model predictions, $\Delta T_{N,M=7}=\sum_{m=1}^{7} t_{N,m}$. Both, measured and predicted curves display a growing magnetic field range of finite \ibal\ as $N$ is increased. It confirms a larger aperture angle of carriers emitted from a QPC at higher modes. Our measured data roughly follow the model curves, albeit they show additional fine structure and a reduced symmetry.

So far we assumed a perfectly flat potential between two perfectly positioned QPCs. The blue solid lines in \fig{fig:magnetic_focusing}c are the result of a more realistic model taking into account the following imperfections of the sample: 
(i) Both QPCs are slightly shifted with respect to each other and the principal axis of the lens, cf. \fig{fig:sample}a. These lateral shifts break the symmetry, such that $I(B)\ne I(-B)$ similar as in our measurements in \fig{fig:magnetic_focusing}b.
(ii) The electrostatic potential beneath the lens is not flat but develops a dip independently of \vl. The dip is caused by the piezoelectric effect of (Al,Ga)As, which is strained by the lens gate during cool-down \cite{supplement}. 
The combination of (i) and (ii) results in additional features in $I(B)$ similar to our experimental observations, albeit the agreement is not perfect: Compared to our model our measurements in \fig{fig:magnetic_focusing}b show for $N\lesssim5$ enhanced transmission for the outermost maxima (at larger $|B|$). This is also visible as an almost bimodal current distribution in \fig{fig:magnetic_focusing}a. We attribute the differences to the scattering properties of the electrostatic potential dip, visualized in Fig.\ 14 in Ref.\ \cite{supplement}. Not knowing its detailed shape we assume a parabolic dip with smooth edges. Compared to our measurements it slightly underestimates the reduction of $I^\text{ball}$ at $B=0$. Such deviations between theory and experiment illustrate our limited knowledge of the exact potential landscape. More accurate predictions might be reached with self-consistent calculations solving the 3D Poisson and Schrödinger equations, which is beyond the scope of this article.

\begin{figure*}[t]
\includegraphics[width=2.1\columnwidth]{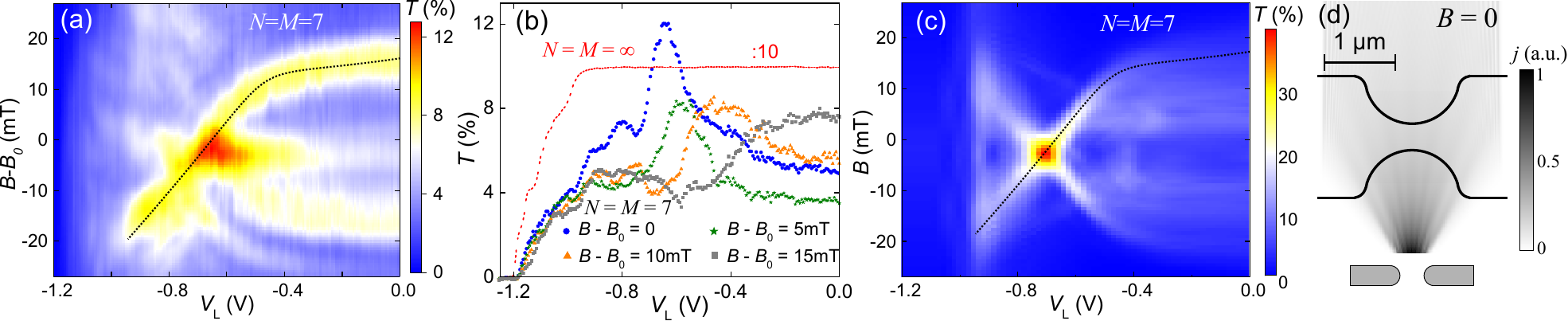}
\vspace{-5mm}
\caption{ (a) Measured serial transmission through both QPCs, $T(\vl,B)$ for $N=M=7$ and QPC$_1$ as emitter. %Dashed line: guide for the eyes.
(b) $T(\vl)$ for various magnetic fields. For $B-B_0=0$ (blue) a pronounced maximum indicates focusing. Red dashed line: $T(\vl)/10$ without QPCs ($V_1=V_2=0$).
(c) Calculated $T(\vl,B)$ as described in the main text. The dashed lines in (a) and (c) are identical. (d) Calculated current density emitted by QPC$_1$ modeled as hard wall potential for $N=7$ at $B=0$ and $\vl=0$ and neglecting the electrostatic potential dip at the lens waist. Solid lines: approximate extension of the lens potential for $\vl\simeq-0.64\,$V.}
\label{fig:combined_focusing}
\end{figure*}

Next we focus on the interference pattern of the transmission curves $\Delta T_{N,M=7}$, which express the lateral coherence in our setup. The $N$ maxima of each fully coherent model curve [red dashed lines in \fig{fig:magnetic_focusing}c] reflect the order of the lateral eigenmodes. 
A classical calculation without disorder \cite{supplement} reproduces the widths and heights of $\Delta T_{N,M}(B)$ in \fig{fig:magnetic_focusing}c but predicts a smooth transmission maximum without oscillations. The dashed gray lines in \fig{fig:magnetic_focusing}c and a copied version in \fig{fig:magnetic_focusing}b are guides to the eyes. They are chosen to connect the $n$th maxima for odd (even) $N$ for the bare model in \fig{fig:magnetic_focusing}c. They also cut through the respective minima for even (odd) $N$, a fingerprint of the coherent mode structure. The measured data in \fig{fig:magnetic_focusing}b approximately reproduce the alternation between minima and maxima found in our model calculations. The comparison confirms the coherent nature of the measured interference pattern. \footnote{Disorder scattering could be considered as an alternative explanation for the observed oscillations. However, to describe our measurements, this would require different arrangements of impurities for each $N,M$, an unrealistic scenario.}

For practical applications it is desirable to maximize the coupling of distant nanodevices, e.g.\ by refocusing carriers emitted from one QPC to the other. To achieve electrostatic focusing, we add a concave spherical lens in the center between the two QPCs, cf.\ \fig{fig:sample}a \cite{Spector1990,Sivan1990}. In a classical model with perfect geometry its focusing properties are described by the electronic version of Snell's law with the refractive index for electrons, $\nr=\sqrt{\ef^0/\efl}$, where the Fermi energies below the lens gate, $\ef^\text{L}$, and elsewhere, $\ef^0$, are assumed to be constants. For our concave lens focusing requires $\nr>1$, i.e.\  $\ef^\text{L}<\ef^0$, which we achieve by applying $\vl<0$.
In \fig{fig:combined_focusing}a we combine electrostatic focusing and magnetic deflection and plot the measured transmission $T(B,\vl)=I(B,\vl)/GV$ for $N=M=7$. (The magnetic deflection experiment shown as gray symbols in \fig{fig:magnetic_focusing}{a}  corresponds to the vertical cross section at $\vl=0$.) While we decrease $\vl<0$ the current maxima bend inwards and eventually cumulate in a single peak at $B=0$ and $\vl\simeq-0.64\,$V, \textit{a direct signature of electrostatic focusing}. Figure \ref{fig:combined_focusing}(b) presents various horizontal cuts $T(\vl)$ for constant $B$. Independent of $B$ the lens pinches off near $\vl=-1.2\,$V similar as the lens' transmission curve without QPCs ($V_1=V_2=0$), added as a red dashed line. Interestingly, the transmission maxima all lie within the range of \vl\ in which the lens itself causes virtually no reflection, corroborating our interpretation in terms of electrostatic focusing. 

While we measure electrostatic focusing as a function of \vl, model calculations are performed in terms of the electrostatic lens potential parameterized by \nr\ or \efl. A direct comparison therefore requires a calibration of $\nr(\vl)$. We combined two complementary methods, namely Landau-level reflection measurements \cite{Taubert2011b} and a self-consistent approach based on Snell's law \cite{supplement}. The calibration allows us to display our model calculations in \fig{fig:combined_focusing}c in the same coordinate system as the measurements in panel (a). The dashed lines in figures \ref{fig:combined_focusing}(a) and (c) are identical and serve as a guide for comparison. The model calculations clearly reproduce the main features of our measurements.

Figure \ref{fig:combined_focusing}(d) shows the calculated current density emitted by a QPC for $N=7$ at $B=0$ into a flat 2DES together with the actual lens geometry. It confirms that the lens captures the emitted beam for $N\le7$, in agreement with the focusing results plotted in \fig{fig:combined_focusing}a. In Ref.\ \cite{supplement} we show that the emission of a QPC depends on the shape of its confinement potential and that a parabolic confinement is in disagreement with our experiment.

For $N=7$ our model predicts a transmission at the focal point of $T\simeq35$\,\%. About half of the reduction from 100\,\% is caused by the discussed imperfections of the layout. The other half is due to an additional lens abberation incorporated by design: we optimized the lens for $N=1$ and thereby neglected the effects of bent electron beams (in contrast to straight beams in ray optics). For a bent beam the lens' focus point depends on the curvature at which carriers are emitted from a QPC. The measured transmission at the focal point is $T\simeq13$\,\%. This further reduction indicates additional deviations of the electrostatic potential from the simulated geometry not yet accounted for in our model.

In summary, using a field effect lens we have achieved electrostatic focusing of ballistic electrons at $B=0$ between two QPCs separated by a mesoscopic region of grounded 2DES. As a tool to directly illustrate electrostatic focusing and to characterize the coherent lateral mode structure of the beam emitted by a QPC, we have combined electrostatic focusing with magnetic deflection. The emission profile of a QPC crucially depends on the shape of its electrostatic (confinement) potential. We present a single-particle quantum-mechanical model which provides realistic predictions of the coherent and ballistic electron dynamics for a given electrostatic potential landscape. 
The quality of its prediction depends on the accurate knowledge of the electrostatic potential. Atomic force and electron beam microscopy allow a precise determination of gate geometries. For the calibration of individual potential components (QPCs and lens) we apply (magneto) transport spectroscopy. Finally, comparison of the measured and calculated current profiles $I(B, V_L)$ through both QPCs in series allows to extract further information on the electrostatic potential landscape such as the dip at the lens waist or details of the QPC confinement potentials.
The accurate description of ballistic electrons will be key for designing future integrated quantum circuits with multiple components. Our results and methods (in experiments and theory) present an important step towards this goal.

We thank Sergey Platonov and Philipp Altpeter for technical support, Sergey Platonov and Yukihiko Takagaki for helpful discussions, and we are grateful for financial support from the DFG via Grant No. LU 819/11-1. MG acknowledges support by project A03 of the CRC-TR 183.

J. Freudenfeld and M. Geier contributed equally to this work.

\begin{center}
\textbf{Supplemental Material\\Coherent Electron Optics with Ballistically Coupled Quantum Point Contacts}
\end{center}

\tableofcontents

\section{Numerical solution of the Schr\"odinger equation}\label{sec:model}

\subsection{Outline of the model}

We model the ballistic transmission of electrons through the two QPCs in series with an open region of grounded 2DES in between by solving the Schr\"odinger equation for a single electron in a 2DES with leads and a potential landscape that implements the QPCs, the lens and the impurities. We treat the 2DES at cryogenic temperatures as a degenerate Fermi gas but take into account energy broadening of the emitted electrons by both, temperature and a finite bias. Supported by experimental results (not shown), we assume coherent transport between the QPCs. However, standing waves between the QPCs due to multiple coherent reflections cannot be resolved because of energy broadening caused by temperature and applied voltage. We model the QPCs using potentials that change slowly on the scale of the Fermi wavelength such that no reflections occur as long as the potential maximum stays below the chemical potential of the 2DES. Our approach is to divide the simulation region into parallel slices of width $\Delta x$ for which we solve the scattering problem exactly. Concatenation of the resulting slice-by-slice scattering matrices in the correct order yields the scattering matrix of the complete simulation region from which we then extract the transmission probabilities $T_{N,M}$ for emitter and detector QPCs on the $N$th and $M$th plateau for a direct comparison with experiments.

\begin{figure}[t]
\includegraphics[width=\columnwidth]{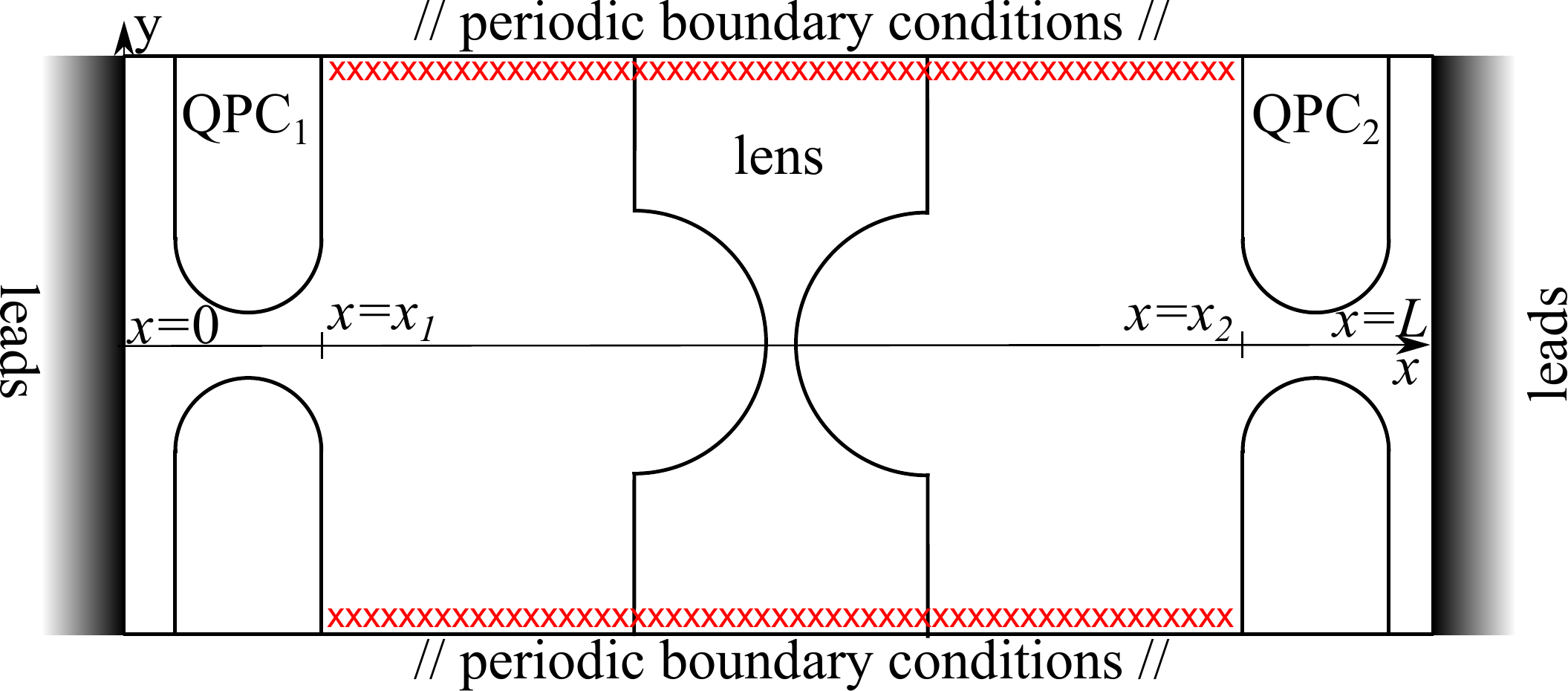}
\caption{Sketch of the simulation region of our model. Gradients on the left and right side denote attached leads. The shapes of two QPCs and a lens which we model by the applied potential $\Phi(x,y)$ are roughly indicated. Our model includes absorbing potentials (red crosses) at the boundaries in y direction to absorb carriers that leave the simulation region.
\label{fig:sketch}}
\end{figure}

The problem of a single electron moving in a magnetic field $B$ and a spatial potential $\Phi(x,y)$ is described by the Hamiltonian
\begin{equation}
H = \frac{1}{2m^\star} \left(- i \hbar \frac{\partial}{\partial \vec{r}} - \frac{e}{c} \vec{A}(x,y)\right)^2 + \Phi(x,y) 
\end{equation}
where $m^\star$ and $e$ are the effective mass and charge of the electron and $c$ is the speed of light. 
While we calculate the solutions at fixed energy $E$, we account for the energy distribution of charge carriers due to finite temperature and source drain voltage by sampling over the energy distribution of the electrons.
The simulation region is sketched in \fig{fig:sketch}{}.

Assuming ballistic transport the potential $\Phi(x,y)$ models two QPCs and a tunable lens in between. We calculate the transmission through QPC$_2$ (detector) for electrons emitted by QPC$_1$ (emitter), i.e., the principal direction of particle flow is along the x-axis. We attach leads to the two boundaries of the simulation region at $x=0$ and $x=L$, each of which is a semi-infinite conducting 2DES without magnetic field. The real sample used in our experiments and shown in Fig. 1(a) of the main text contains four additional ohmic contacts arranged symmetrically along the sides to ground the 2DES region between the two QPCs. These macroscopic and diffusive contacts absorb electrons scattered to the sides such that they cannot escape through one of the two QPCs. We simulate the side contacts by an absorbing potential along the boundaries of the simulation region in $y$-direction together with periodic boundary conditions in $y$-direction which allow an expansion in terms of plane waves. Along the $x$-axis we divide the simulation region into exactly solvable vertical slices (parallel to the $y$-axis) by approximating the potential
\begin{equation}
\Phi(x,y) \simeq \sum_i \Phi(x_i,y) \delta(x - x_i)
\label{eq:qm_phi}
\end{equation}
where we choose a normalization of the Dirac delta function such that $\int_{-\infty}^\infty \text{d}x \delta(x) = \Delta x$, where $\Delta x = x_{i+1} - x_i$. The normalization is chosen to preserve $\int_{x=0}^L \text{d}x \Phi(x,y)$, where $L$ is the length of the simulation region in $x$ direction, compare \fig{fig:sketch}{}. We include a magnetic field $\vec{B}(x) = B(x)\hat{e}_z$ perpendicular to the 2DES which is approximated by
\begin{equation}
B(x) \simeq \sum_i B(x_i) \delta(x - x_i) .
\label{eq:qm_B}
\end{equation}
We choose the corresponding vector potential as $\vec{A}(x) = A_y(x) \hat{e}_y$, where $A_y = \int_0^x \text{d}x'B(x') = \sum_{x_i < x} B(x_i) \Delta x$.
The magnetic field is constant within the region between the QPCs but vanishes in the leads and in the regions with finite QPC potential at $0 \leq x \leq x_1$ and $x_2 \leq x \leq L$. Neglecting the magnetic field within the QPC regions is justified as the cyclotron radius is much larger than the QPC constriction size. To avoid reflections due to discontinuities of $B$ at $x=x_1$ and $x = x_2$, we turn on the magnetic field slowly on the scale of the Fermi wavelength. The solution converges when the slide width $\Delta x$ is much smaller than the Fermi wavelength.

\subsection{Scattering problem in a nut-shell}

\begin{figure}[t]
\includegraphics[width=\columnwidth]{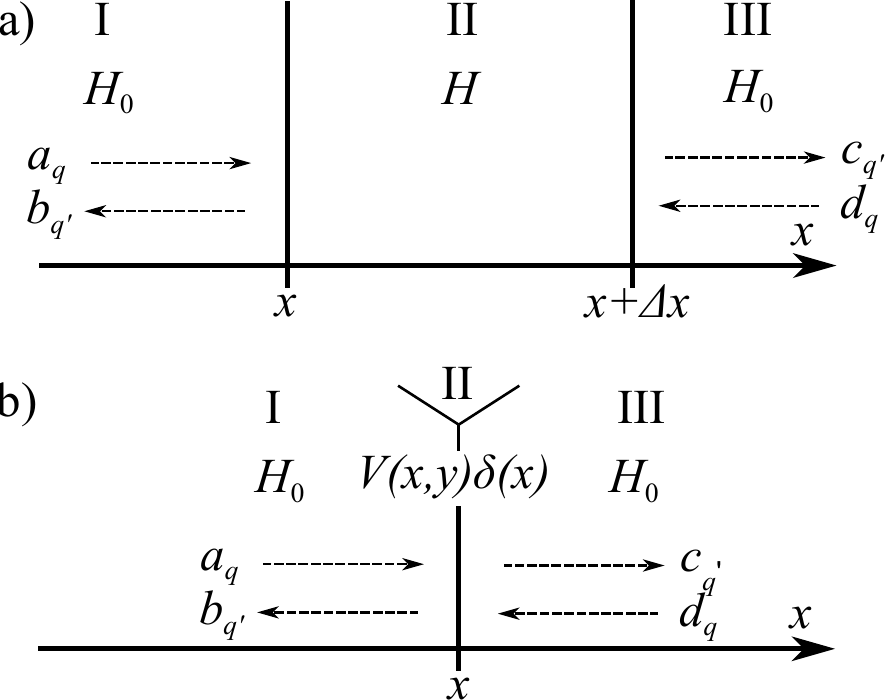}
\caption{Sketch of the model used to compute the scattering matrix per slice. We compose the scattering matrix from two contributions: a) the propagative motion described by the Hamiltonian $H$ defined in \eq{eq:propagation} between two scattering slices at $x$ and $x+\Delta x$ and b) the scattering off the potential $\Phi(x,y)$ per slice at each $x$. The regions I and III are the semi infinite regions with Hamiltonian $H_0 = - \frac{\hbar^2}{2m^\star} \frac{\partial^2}{\partial x^2}$.
\label{fig:sketch_scattering}}
\end{figure}
\begin{figure}[ht]
\includegraphics[width=\columnwidth]{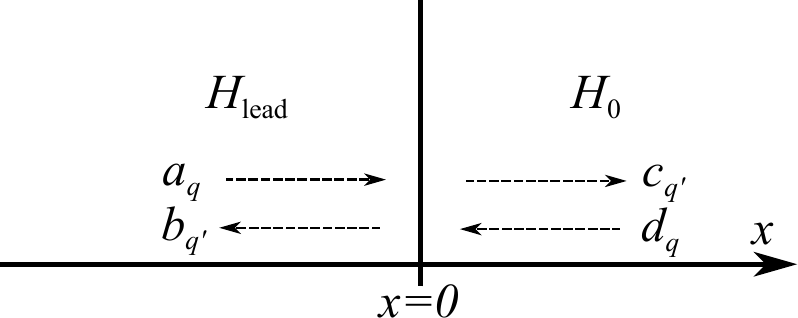}
\caption{Sketch of the model used to compute the scattering matrix of the attached left lead. The lead Hamiltonian $H_{\text{lead}}$ as given in equation \eqref{eq:Hlead} contains only kinetic energy terms. The right region is described by the Hamiltonian $H_0$. For the right leads the setup is mirrored and the boundary between the regions is at $x=L$.
\label{fig:sketch_scattering_leads}}
\end{figure}

For solving the scattering problem, between $x=0$ and $x=L$ we expand the wavefunction along the vertical slices (in $y$-direction) in plane waves characterized by the transverse wavenumber $q$. 
Making use of the linearity of the scattering problem we calculate the scattering matrix of each individual slice separately. For each slice we further divide the problem into a free propagation 
between $x_i$ and $x_{i + 1}$ setting $\Phi = 0$ and $B = 0$, and the actual scattering events at $x_i$ accounting for the potential and the magnetic field as defined in equations \eqref{eq:qm_phi} and \eqref{eq:qm_B}, respectively.
The ``propagative motion'' is sketched in \fig{fig:sketch_scattering}a and the ``scattering events'' in \fig{fig:sketch_scattering}b. Regions I and III are temporarily added supplementary semi-infinite regions described by the trivial Hamiltonian
\begin{equation}
H_0 = - \frac{\hbar^2}{2m^\star} \frac{\partial^2}{\partial x^2}\, .
\label{eq:H0}
\end{equation}
These supplementary leads are completely removed in the last step, where we concatenate the slices in the order of appearance from left to right. Before the final concatenation process we add the semi-infinite leads for $x<0$ and $x>L$ to the system. For the region with $B=0$, the scattering matrix of the uniform leads is sketched in \fig{fig:sketch_scattering_leads}{}.

\subsection{Details of the calculation}\label{sec:matrices}

The scattering matrix $S_{q', q}$ is the unitary matrix that relates the amplitudes of ingoing waves $a_q, d_q$ to the scattered amplitudes of outgoing waves $b_{q'}, c_{q'}$ left and right of the scattering region. 
The scattering event may change the wavenumber of the ingoing wave from $q$ to $q'$. 
The scattering matrix is defined as
\begin{align}
\begin{pmatrix}
b_{q'} \\
c_{q'}
\end{pmatrix} & = \int_{-\infty}^\infty \text{d}q S_{q', q} \begin{pmatrix}
a_{q} \\
d_{q}
\end{pmatrix} \\ 
& = \int_{-\infty}^\infty \text{d}q \begin{pmatrix}
t_{q', q} & r'_{q', q} \\
r_{q', q} & t'_{q', q}
\end{pmatrix} \begin{pmatrix}
a_{q} \\
d_{q}
\end{pmatrix}
\end{align}
where $t_{q', q}$ ($t'_{q', q}$) and $r_{q', q}$ ($r'_{q', q}$) are transmission and reflection matrices for waves incident from the left (right), respectively. 

The scattering matrix for the propagative slices is solved in a model as sketched in \fig{fig:sketch_scattering}a. The Hamiltonian $H$ in region II implementing the propagation of the plane wave eigenmodes from $x_i$ to $x_{i+1}$ is 
\begin{equation}\label{eq:propagation}
H =  -\frac{\hbar^2}{2m^\star} \frac{\partial^2}{\partial x^2} + \frac{(\hbar q - \frac{e}{c} A_y(x_i))^2}{2m^\star} .
\end{equation}
where the term $\frac{e}{c} A_y(x_i)$ describes the piecewise constant vector potential corresponding to the magnetic field as defined in eq. \eqref{eq:qm_B}. For this geometry, the transmission matrices read
\begin{equation}
t_{q', q}  = \frac{2 k k_x e^{- i k \Delta x} \delta(q - q')}{2 \cos \left(k_x \Delta x \right) k k_x -i \sin \left(k_x \Delta x \right) \left( k^2 + k_x^2 \right)}
\end{equation}
and $t'_{q', q} = t_{q', q} e^{2 i k \Delta x}$ and the reflection matrices are
\begin{equation}
r_{q', q} = r'_{q', q} = \frac{i \sin \left( k_x \Delta x \right) \left( k_x^2 - k^2 \right) \delta(q - q')}{2 \cos \left(k_x \Delta x \right) k k_x -i \sin \left(k_x \Delta x \right) \left( k^2 + k_x^2 \right)}
\end{equation}\\
with the wavenumber along $x$ in regions I and III $k = \sqrt{2 m^\star E / \hbar^2}$ and the wavenumber along $x$ in region II $k_x = \sqrt{2 m^\star E / \hbar^2 - (q - \frac{e}{\hbar c} A_y(x))^2}$. The scattering matrix is diagonal in the wavenumber $q$. In the propagative region with $\Phi(x,y) = 0$ they satisfy $k^2 = k_x^2 + q^2$.
To implement the potential $\Phi(x,y)$, we solve the scattering problem for a slice $\Phi(x_i, y) \delta (x-x_i) $ in the geometry depicted in \fig{fig:sketch_scattering} b). The transmission and reflection matrices are
\begin{equation}
t = t' = 2 \left( 2 \id + \frac{2 i m^\star \Delta x}{\hbar^2 k} \mathbf{\Phi} (x_i) \right)^{-1}
\end{equation}
\begin{equation}
r = r' = \left( 2 \id + \frac{2 i m^\star \Delta x}{\hbar^2 k} \mathbf{\Phi} (x_i)\right)^{-1} \cdot \left( \frac{- 2 i m^\star \Delta x}{\hbar^2 k} \mathbf{\Phi} (x_i) \right)
\end{equation}\\
where $\id$ is the identity matrix and $\mathbf{\Phi}_{q_1, q_2}(x_i) = \Phi_{q_1 - q_2}(x_i)$ is the Toeplitz matrix formed by the Fourier transform of the potential $\Phi_q(x_i) = \int_W \text{d}y \Phi(x_i, y) e^{- i q y}$ with $W$ the width of the simulation region along $y$. 
For the leads at the left and right boundary of the simulation region, we assume that there is no magnetic field. The vector potential in the left (right) leads is equal to the vector potential at the left (right) boundary of the simulation region $A_y^\text{lead} = A_y(0)$ ($A_y^\text{lead} = A_y(L)$). 
The Hamiltonian in the leads is
\begin{equation}
H_\text{lead} =  -\frac{\hbar^2}{2m^\star} \frac{\partial^2}{\partial x^2} + \frac{(\hbar q - \frac{e}{c} A_y^\text{lead})^2}{2m^\star} .
\label{eq:Hlead}
\end{equation}
In a scattering geometry for a left lead as in Fig. \ref{fig:sketch_scattering_leads}, the transmission amplitudes read
\begin{equation}
t_{q', q} = t'_{q', q} =  \frac{2 \sqrt{k_x k} }{k + k_x} \delta(q - q')
\end{equation}
if $E \geq (\hbar q - \frac{e}{c} A_y(0))^2 / 2 m^\star$ and $t_{q', q} = t'_{q', q} = 0$ else. 
The reflection amplitudes read 
\begin{equation}
r_{q', q} = - r'_{q', q} = \frac{k_x - k }{k + k_x} \delta(q - q').
\end{equation}
For the leads right of the scattering region, the constant vector potential is $A_y^\text{lead} = A_y(L)$ and $r$ and $r'$ are exchanged. 

The concatenated scattering matrix $S_{12}$ of two adjacent regions 1, 2 with 1 left of 2 can be computed from the individual scattering matrices $S_1$, $S_2$ as 
\begin{align}
t_{12} & = t_2  \left(\id - r'_1 r_2\right)^{-1}  t_1 \\
r_{12} & = r_1 + t'_1  r_2 (\id - r'_1 r_2)^{-1}  t_1 \\
t'_{12} & = t'_1  \left(\id + r_2 (\id - r'_1  r_2)^{-1}  r'_1\right)  t'_2 \\
r'_{12} & = r'_2 + t_2 \left(\id - r'_1  r_2\right)^{-1} r'_1  t'_2 .
\end{align}
To compose the scattering matrix of the complete system, we concatenate all propagating and potential scattering matrices ordered from left to right and complete the calculation by concatenating with the lead scattering matrices.

\subsection{Transmission matrix}

\begin{figure}
\includegraphics[width=1.0\columnwidth]{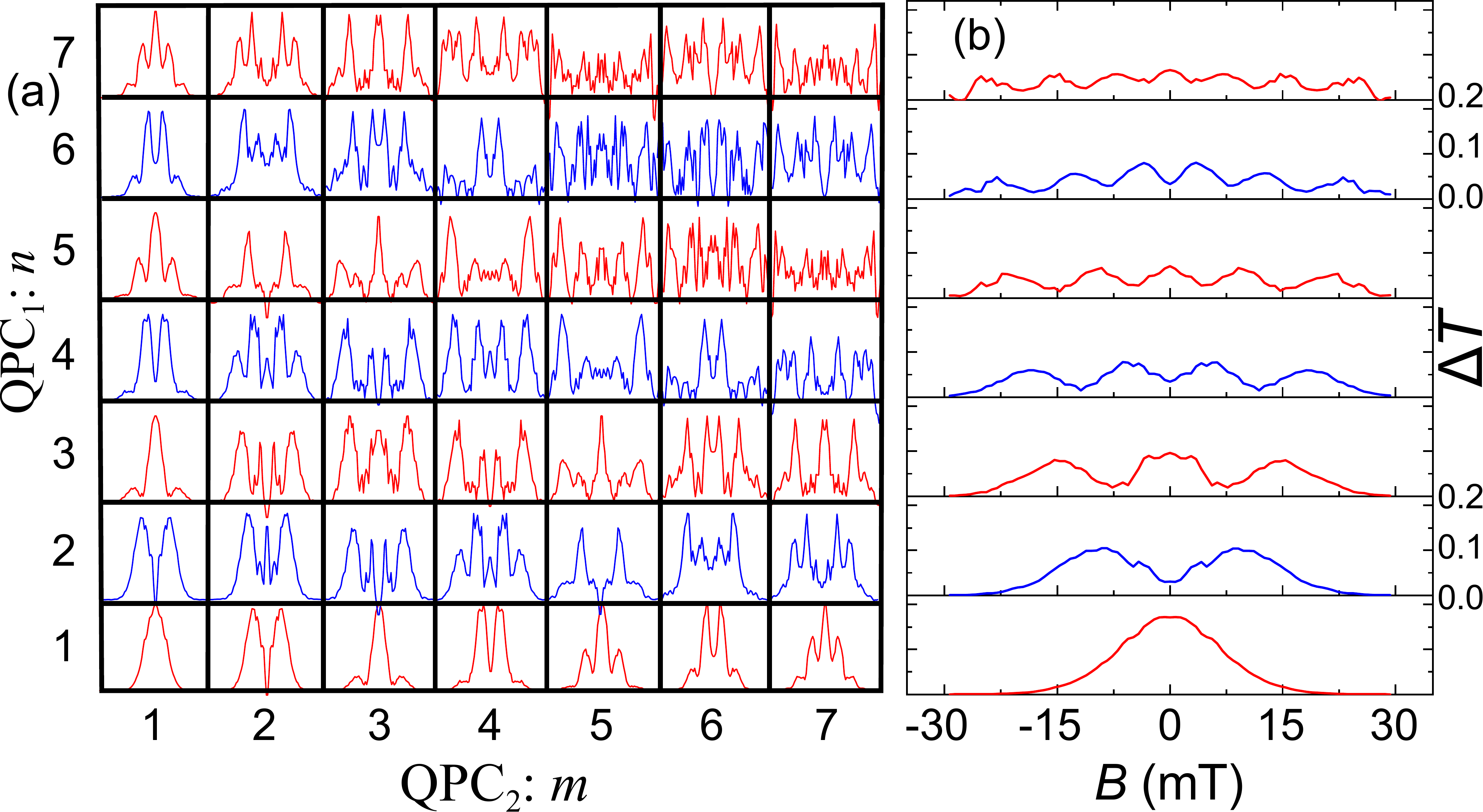}
%\vspace{-3mm}
\caption{
(a) Calculated transmission probabilities ---for $\vl=0$ (no lens) and a flat electrostatic potential between the perfectly aligned QPCs --- averaged over energy to simulate temperature and source-drain voltage broadening. Each square shows $t_{n,m}(B)$ between the $n$th mode of QPC$_1$ and the $m$th mode of QPC$_2$ for $1\le n,m\le 7$. $B=0$ at vertical symmetry axes. For $N=M=7$ the total transmission is the sum of all shown elements $T_{N,M}=\sum_{n,m=1}^{N,M}t_{n,m}$.
(b) Transmission differences $\Delta T_{N,M=7}=\sum_{m=1}^{7} t_{N,m}$ also plotted in Fig.\ 2(c) of the main article. They correspond to the sum of the colored elements in each row of the matrix in panel a.} 
\label{fig:transmission}
\end{figure}

In the main article we expressed the total transmission probability in terms of a transmission matrix $T_{N ,M} = \sum_{n, m}^{N, M} t_{n, m}$ for emitter and detector QPC on the $N$th and $M$th conductance plateaus. Here, $t_{n, m}$ are the individual probabilities for an electron emitted from the $n$th mode to transmit through the $m$th subband of the detector. We present them in \fig{fig:transmission}a.
This approach is exact as long as the eigenmodes do not depend on $N$ and $M$ and coherent reflections between the QPCs can be neglected. The latter is fulfilled in our experiments as the dephasing length is small compared to the distance between the QPCs $l_\phi\simeq0.5\,\mu\text{m}\ll l$. However, in Ref.\ \cite{Geier2020-1} we show that the lateral confinement potential of the QPCs is modified by Coulomb screening as the mode number increases. As a consequence the eigenmodes are also affected and the transmissions $t_{n, m}$ depend on $N$ or $M$, respectively. The descriptive picture presented in the main article is a good approximation as long as the change of the confinement potential between subsequent plateaus of the QPCs is small.  

While the approximation works quite well for our experiments, in our numerical calculations described in the sections above we nevertheless directly calculate the total transmission $T_{N ,M}$ for the realistic electrostatic potential which we determined from the experimental subband spacings \cite{Geier2020-1}.

For completeness in \fig{fig:transmission}b we present the transmission differences $\Delta T_{N,M=7}=\sum_{n=1}^{7}t_{n,M=7}$ each corresponding to summing up all elements in a row of the matrix in panel (a). These transmission differences are also discussed in the main article, cf.\ Fig.\ 2 of the main article. 

\subsection{Analytic description of QPC potentials}
\label{sec:QPC potentials in the numerical simulations}

In a separate work in Ref.~\onlinecite{Geier2020-1} we show that the confinement potential at the center of our gate defined QPCs on the $N \leq 2$ conductance plateau is well described by a parabola while for $N \geq 3$ it is best described by a hard-wall potential. 
Below we introduce a two dimensional continuation of both the hard-wall and parabolic potential. 
We assume that the magnetic field can be neglected within the QPC regions in both quantum and classical simulations. This is a good approximation as long as the magnetic length is large compared to the channel width at the QPCs, as in all our measurements presented here. 

\begin{figure}
\includegraphics[width=0.9\columnwidth]{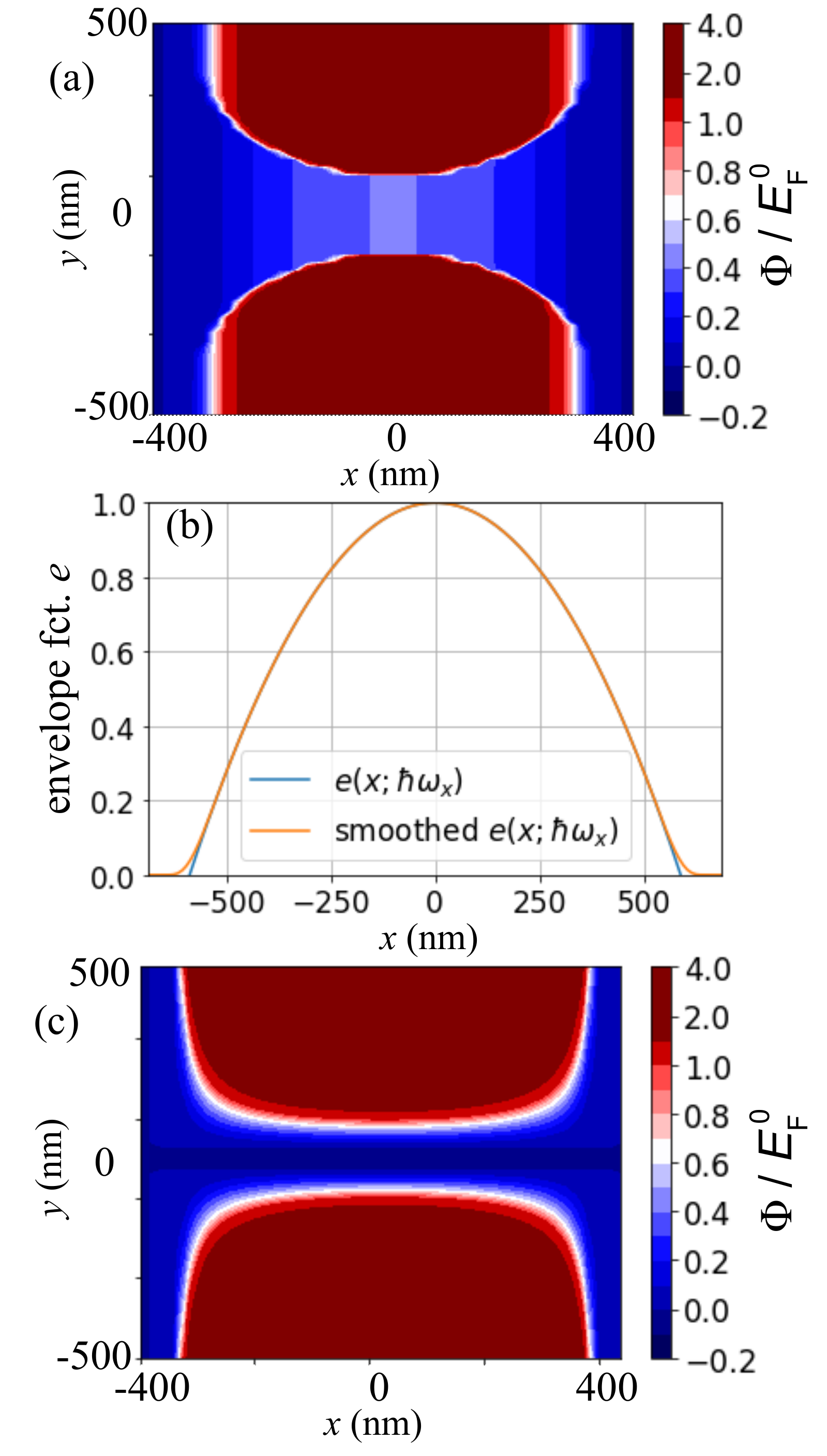}
\caption{(a) Hard-wall potential modeling a QPC on the $N = 7$th conductance plateau with parameters from table~\ref{tab:QPC_Parameters}.  
(b) Envelope function $e(x; \omega_x)$ as defined in eq. \eqref{eq:envelope} (blue) and the smoothed version used in the quantum mechanical simulation (orange).
(c) Smooth potential modeling a parabolic QPC on the $N = 7$'th conductance plateau. 
%The values closely resemble our QPCs (cf.\ Table \ref{tab:QPCfitreport} in chapter \ref{sec:QPC-parabolicity}) and we use them in our simulations. We use a slightly smaller value for $\hbar \omega_x$ accounting for deviations from a perfectly parabolic current barrier at the transition between the QPCs and the free 2DES.
}
\label{fig:potentials}
\end{figure}

\textit{Hard-wall potential.} 
The electrostatic potential realizing the QPCs based on a hard-wall model can be defined within the QPC regions $0 \leq x \leq x_1$ and $x_2 \leq x \leq L$ as 
\begin{equation}
\phi(x,y)=  \begin{cases}
 e(x, \omega_x) \phi_0 & |y| \leq \frac{W}{2} + \sqrt{(R_0 - \frac{W}{2})^2 - x^2} \\
 4 e(x, \omega_x) E_\text{F}^0 & \text{else}
\end{cases} 
\label{eq:classical_slit}
\end{equation}
where $e(x, \omega_x)$ is an envelope function as described below and $E_\text{F}^0$ is the Fermi energy of the free 2DES, i.e. the kinetic energy of electrons in the free 2DES at the Fermi edge.
The electrostatic potential \eqref{eq:classical_slit} has a discontinuity along two semicircles modeling the semicircular shape of the split gates defining the QPC. The width $W$ and the offset potential $\phi_0$ at the QPC center determines the mode structure of the QPC. These parameters can be determined from source-drain bias voltage spectroscopy \cite{Geier2020-1} and the result is summarized in table \ref{tab:QPC_Parameters}.
\begin{table}[h]
  \centering
	\begin{tabular}{ |c|c|c|c|c|c|c|c| } 
 \hline
 $N$ & 1 & 2 & 3 & 4 & 5 & 6 & 7  \\ 
  \hline
 $W$\,(nm) & 58 & 84 & 111 & 134 & 160 & 193 & 221 \\ 
  \hline
 $\Phi_0 / \ef^0$ & 0.62 & 0.52 & 0.47 & 0.41 & 0.38 & 0.41 & 0.41 \\  
 \hline 
	\end{tabular}
\caption{Quantum well width $W$ and offset potential relative to the Fermi energy $\Phi_0/ \ef^0$  for a hard-wall confinement potential determined from source-drain bias spectroscopy in Ref.~\cite{Geier2020-1}}
\label{tab:QPC_Parameters}
\end{table}
The radius $R = R_0 - W/2$ of the semicircles decreases with increasing width of the QPC center modeling the shrinking of the gate depleted region when the gate voltage is increased.
In our simulations we set $R_0 = 400$\,nm which yields a slightly larger distance between the centers of the semicircles than in the QPC gate geometry ($d/2 = 275$\,nm). The value of $R_0$ is chosen in order to find better agreement in the width of the transmission profile $T(B)$ with the magnetic deflection experiments. The correction accounts for uncertainties in the precise determination of the electrostatic QPC potential.

In current direction we assume a parabolic transition of the electrostatic potential between the 1D channel and the free 2DES in order to ensure a reflectionless transmission through the QPCs. This is implemented by a smooth version of the envelope function
\begin{equation}
e(x; \omega_x) = \begin{cases}
 0, & |x| \geq \frac{L_\text{QPC}}{2} \\
 1 - \frac{m \omega_x^2 x^2}{2 E_\text{F}^0}, & \text{else}
\end{cases} 
\label{eq:envelope}
\end{equation}
where the channel length $L_\text{QPC} = \frac{2}{\omega_x} \sqrt{\frac{2 E_\text{F}^0}{m}} = 2 R_0 - W$ such that the function $e(x, \omega_x)$ is continuous. The potential as well as the envelope function are plotted in \fig{fig:potentials}{}. In the quantum mechanical simulation, we use a version of $e(x, \omega_x)$ that connects smoothly to $\Phi(x > x_1, y) = 0$ over a length $L_\text{QPC} / 12$ on both sides of the QPC as shown in \fig{fig:potentials}{b}. 
%In the classical simulation we include a region of length $L_\text{QPC} / 12$ at the boundaries of each QPC with vanishing potential and magnetic field to substitute the smoothing region in the quantum mechanical simulation. 
%To simulate a QPC at the $N$'th conductance plateau we set $\Phi_0 = \mu - n \hbar \omega_y$.

\textit{Parabolic saddle point potential.} 
For comparison, a potential that has a parabolic saddle point at the QPC center and connects continuously to the free regions with vanishing potential can be defined as
\begin{equation}
\Phi(x,y) = e(x; \omega_x) \left(\Phi_0 + \frac{m \omega_y^2 y^2}{2} \right) .
\label{eq:QPCpotential_sim}
\end{equation}
where $\hbar \omega_y$ is the subband spacing. To simulate a QPC at the $N$'th conductance plateau we set $\Phi_0 = E_\text{F}^0 - n \hbar \omega_y$.

\subsection{Implementation of the lens potential in our numerical model}

\begin{figure}[tbh]
\includegraphics[width=0.8\columnwidth]{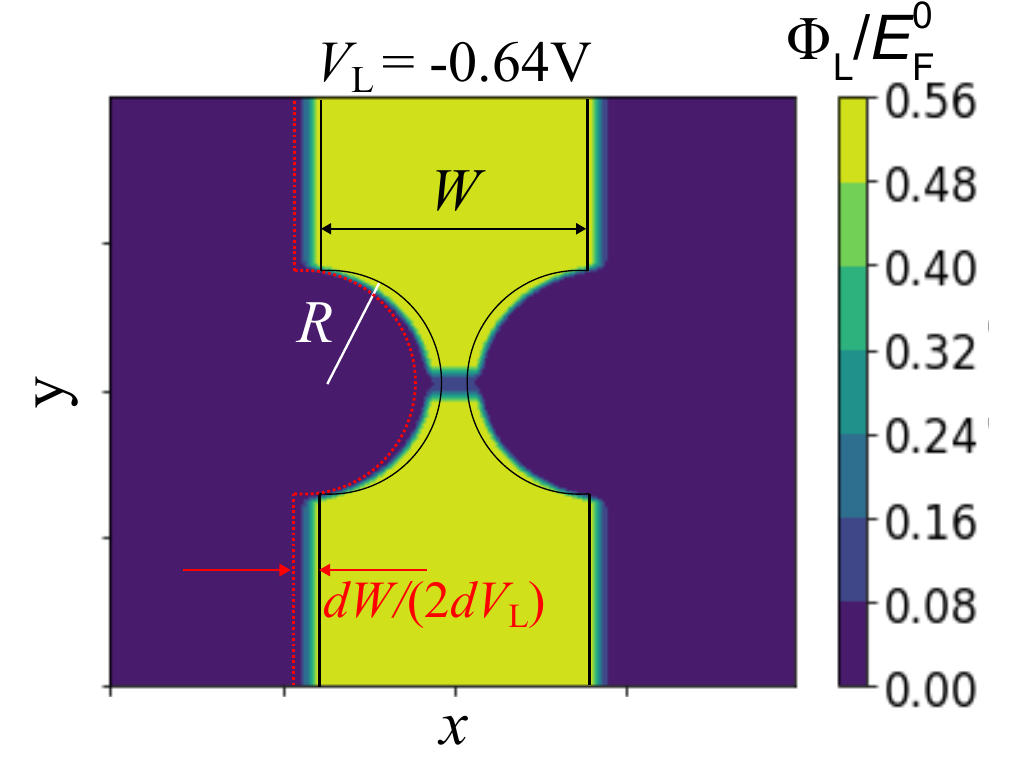}
\caption{Model implementation of the lens potential surrounded by free 2DES at zero potential (blue). The lens gate voltage $\vl=-0.64$\,V corresponds to the focusing condition.
} 
\label{fig:lenspotential-implementation}
\end{figure}

In \fig{fig:lenspotential-implementation}{}
we illustrate the lens potential implemented in our numerical model, accounting for our lens calibration described in detail in Sec. \ref{sec:calibration} below. The lens represents a region of finite potential in the otherwise free 2DES. The region of finite potential mimics the shape of the lens gate, cf. Fig. 1(a) in the main paper, with the width $W=1780$\,nm and lens radius $R=750$\,nm. The potential width is additionally increased by $\Delta W / \Delta \vl = 75$\,nm/V, taking into account the measured depletion length caused by the electric field effect. A dip in the center of the lens is caused by a piezoelectric potential, described in Sec. \ref{sec:calibration} below. To avoid unrealistic sharp edges of the lens potential we model the potential transitions by smooth step functions.

\section{Comparison between quantum mechanical and classical solution}
\label{sect:classicalModel}

In order to identify coherence effects we compare our results with a classical simulation of the magnetic deflection experiment. To compute the current for classical electrons we sample the lateral distribution function at the center (in current direction) of the emitter. We solve the phase space trajectory of each sampling point numerically and determine whether it transmits through the detector and hence contributes to the current. In the classical simulation we neglect trajectories that involve multiple reflections between the QPCs. 

\begin{figure}
\includegraphics[width=\columnwidth]{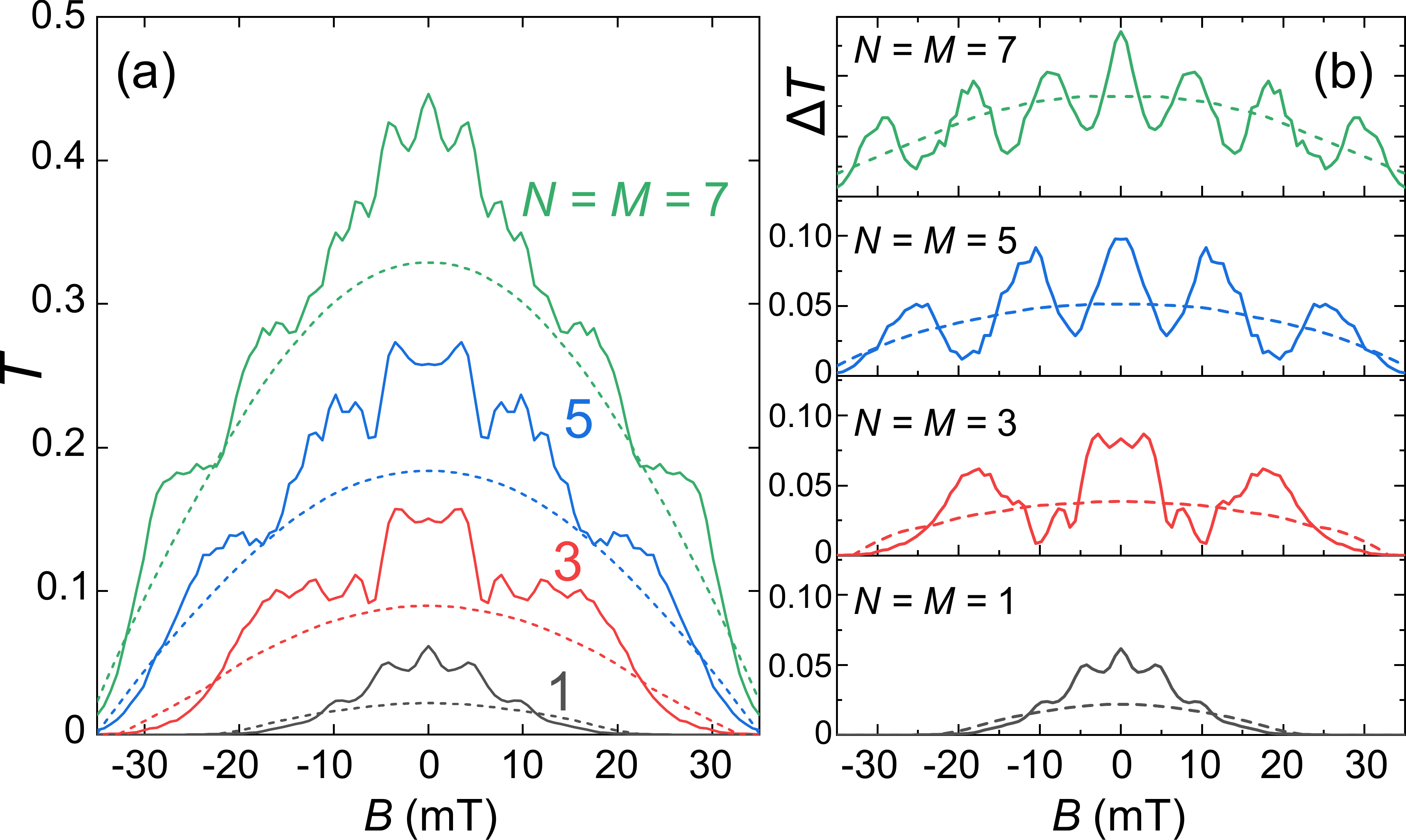}
\caption{Comparison of quantum mechanically (full lines) and classically (dashed lines) transmissions through the two QPCs in series calculated for the harmonic electrostatic QPC potential in the perfectly aligned geometry and without the electrostatic potential dip at the lens waist caused by the piezoelectric effect. (a) $T_{N,M}(B)$ and (b) corresponding first transmission differences $\Delta T_{N,M}(B)=T_{N,M}(B)-T_{N-1,M-1}(B)$. Both emitter and detector QPC are set to the same conductance plateau $N=M=1,3,5,7$.
\label{fig:QM_vs_Classical}}
\end{figure}

In \fig{fig:QM_vs_Classical}{a}
we show the classically and quantum mechanically computed magnetic deflection transmission profiles $T(B)$ for a perfectly aligned geometry and without the electrostatic potential dip at the lens waist. Both computations were performed with identical parameters using the harmonic QPC potential. The quantum mechanical solution exhibits oscillations that reflect the mode structure of the waves emitted from the QPC. These coherent oscillations are absent in the transmission profile for classical electrons. 

The envelope of the transmission profile of coherent electrons coincides with the classical result for large magnetic field strengths. For small magnetic fields there are electron trajectories involving reflections between the QPCs that contribute to the current through the setup. These trajectories are included in the quantum mechanical calculation but are neglected in the classical calculation. They yield an overall enhanced transmission for small magnetic field strengths in the quantum mechanical calculation compared to our classical solution. 

In our experiments coherent Fabry-Perot like oscillations caused by standing waves between the QPCs are averaged out, as expected for the dephasing length of $l_\phi = 500\,$nm due to bias and temperature broadening, cf.\ Fig. 2 of the main article. In our quantum mechanical calculation we account for the broadening by energy averaging. Nevertheless, the quantum mechanical solution in \fig{fig:QM_vs_Classical}{a} contains high frequency oscillations at small magnetic fields which are a remnant of the Fabry-Perot oscillations after averaging over the energy distribution of transmitting electrons. These remnants of the Fabry-Perot oscillations are vulnerable to geometric imperfections and weak disorder contributing additional phase shifts to interfering paths. Therefore they are most pronounced in the perfectly aligned geometry without electrostatic potential dip at the lens waist.

In \fig{fig:QM_vs_Classical}{b} we present the calculated first transmission differences $\Delta T_{N,M}(B)=T_{N,M}(B)-T_{N-1,M-1}(B)$, again using $N=M$. As for $T(B)$ in panel (a), the classical solution is a smooth maximum, while the quantum mechanical result oscillates strongly with $N=M$ main maxima reflecting the QPC mode structure.

The lateral mode structure of the QPCs is visible in both, experimental results and quantum mechanical solution of their serial transmission, cf.\ Fig. 2 of the main article. In contrast to the Fabry-Perot like oscillations discussed above the lateral mode structure remains stable as long as the dephasing length is larger than the lateral distance of two adjacent maxima in the interference pattern of emitted electrons.

\section{Emission profiles for hard-wall versus parabolic lateral potentials}

In this section we compare the calculated emission profiles for our two models introduced in section \ref{sec:QPC potentials in the numerical simulations} describing the electrostatic QPC potential. 

\subsection{QPC emission profiles}

\begin{figure}[htb]
\includegraphics[width=1.0\columnwidth]{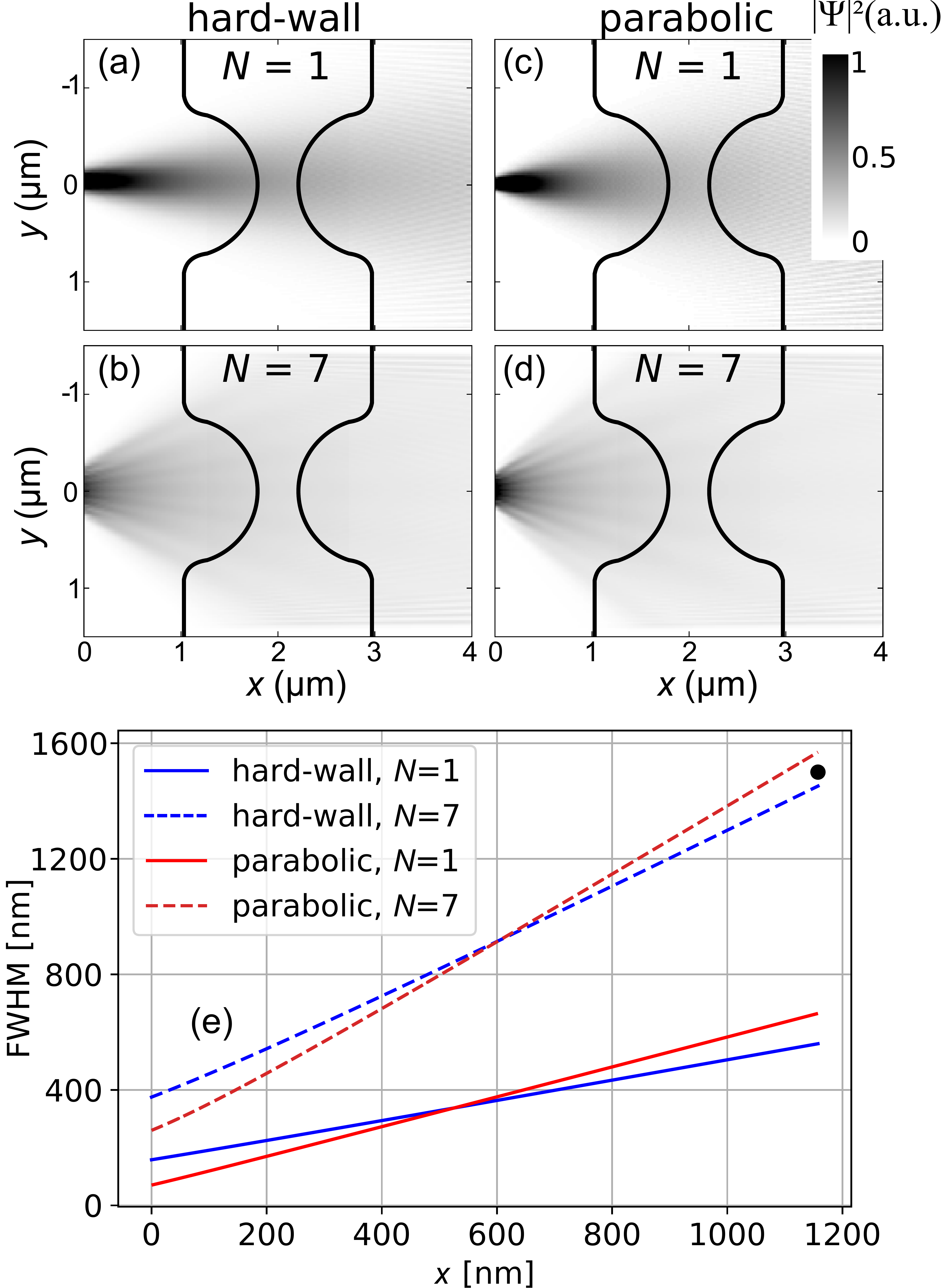}
\caption{Beam profiles of electrons emitted from a QPC with hard-wall potential a), b) and a parabolic potential c), d) as defined in section \ref{sec:QPC potentials in the numerical simulations}. In a), c), the QPCs are set to the first conductance plateau. In b), d), the QPCs are set to the seventh conductance plateau. We include a sketch of the lens gate accounting for a widening of the lens potential of around $100$\,nm for the voltage at the focus point. Figure e) shows the extracted full width half maximum (FWHM) of the beams as a function of the position $x$ in the free region between the QPCs. The black dot denotes the lens aperture and position. 
} 
\label{fig:beamprofiles}
\end{figure}

The beam profiles are calculated from the solution of the Schr\"odinger equation, namely the spatially resolved probability density of an electron emitted from the QPC at the Fermi energy. In \fig{fig:beamprofiles}{}
we compare the beam profiles for the hard-wall versus parabolic potential. The beam profiles of the hard-wall potential have a wider wist and smaller divergence compared to the parabolic potential. This tendency is additionally presented in \fig{fig:beamprofiles}e by plotting the full-width-half-maximum (FWHM) of the respective beams as a function of the distance to the QPC.

\subsection{Magnetic deflection and electrostatic focusing properties}

\begin{figure}[htb]
\includegraphics[width=0.9\columnwidth]{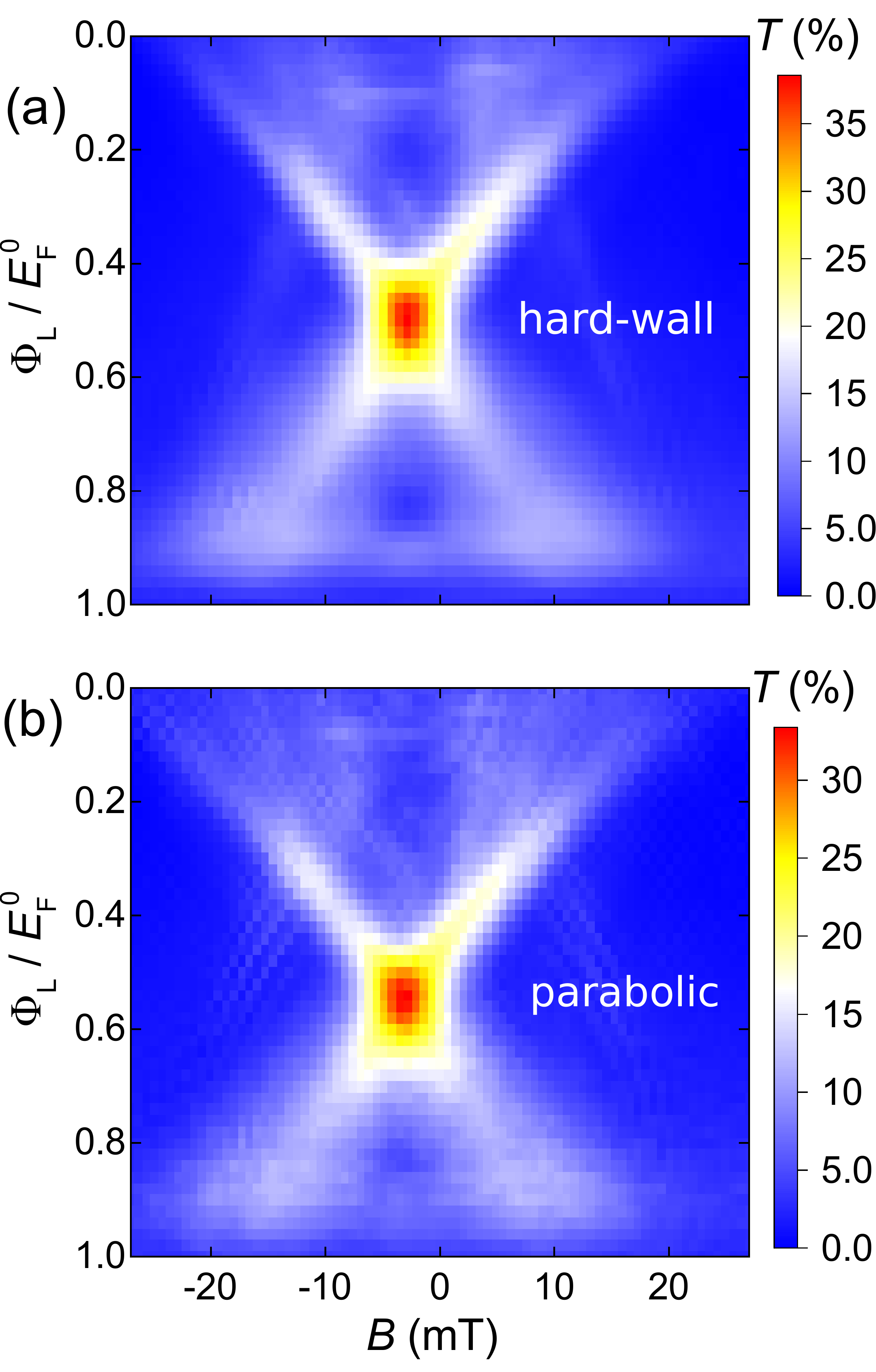}
\caption{Simulated magnetic deflection transmission probabilities $T(\Phi_\text{L}, B)$ with finite electrostatic lens potential $\Phi_\text{L}$ for emitter and detector QPC set to $N = M = 7$ conducting channels with (a) hard-wall QPC potential model and (b) parabolic QPC potential model.} 
\label{fig:hardwall_vs_parabola_TVLB}
\end{figure}

In \fig{fig:hardwall_vs_parabola_TVLB}{}
we compare the combined magnetic deflection with electrostatic focusing, namely the transmission probabilty through the two QPCs for $B\ne0$ and $\vl\ne0$, assuming a hard-wall (panel a) versus parabolic QPC (b) potential for $N = M = 7$. Both QPC models show electrostatic focusing. The main differences between the two models is that the parabolic QPC potential exhibits a broader transmission profile at vanishing lens potential and a reduced transmission at the focal point. Both differences are caused by the larger divergence of the parabolic QPC potential. The larger divergence of the QPC potential implies that not all electrons can be captured by the lens, cf.\ \fig{fig:beamprofiles}{}, and therefore a reduced transmission at the focal point. 

From our experimental data we concluded that all electrons are captured by the lens, in contradiction to the beam profile of the parabolic QPC potential. Furthermore, the measured magnetic field dependence of the transmission profile at $\vl=0$ is in better agreement with the predictions of the hard-wall model than those of the parabolic confinement model. 

\begin{figure*}[ht]
\includegraphics[width=2.0\columnwidth]{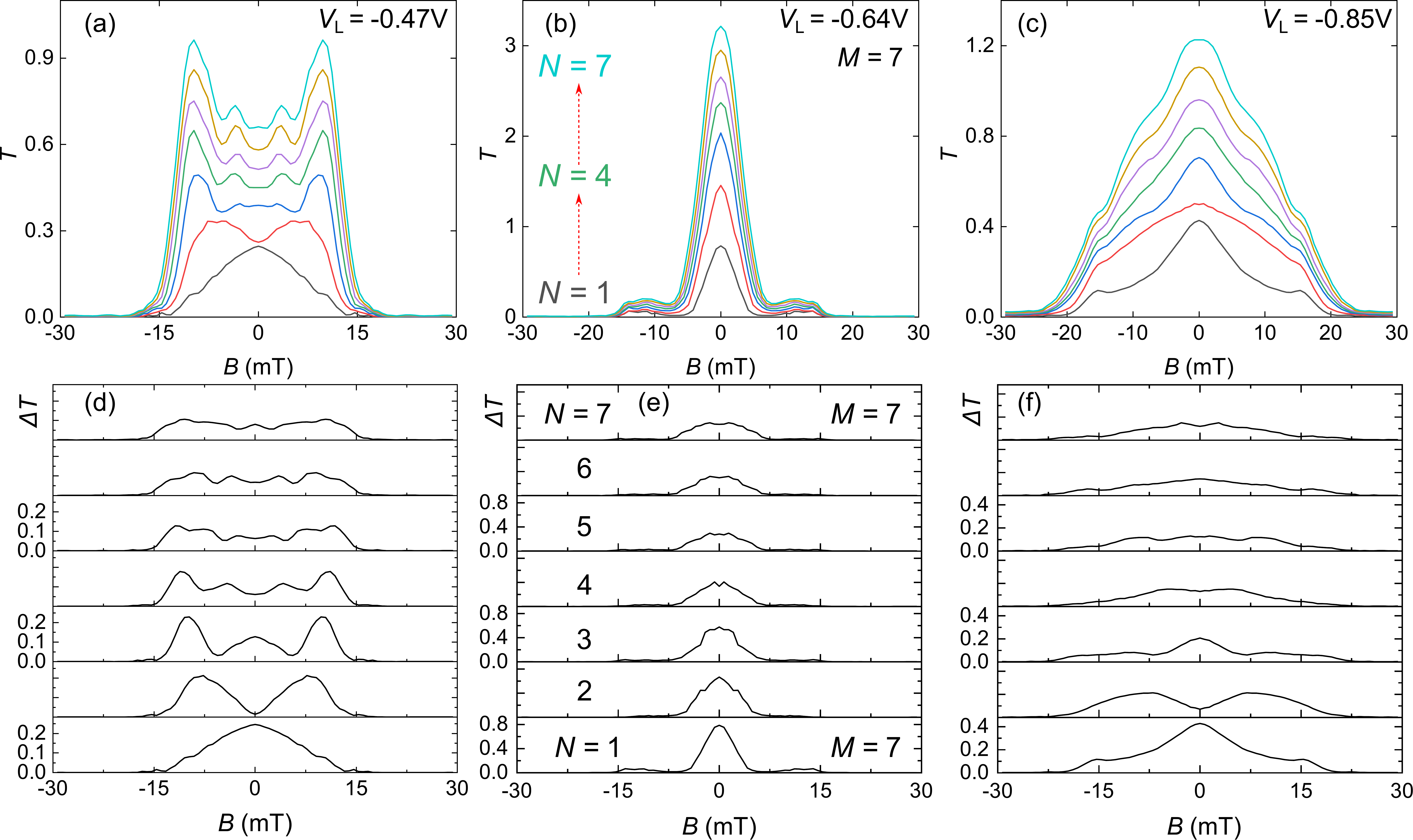}
\caption{Calculated transmissions $T(B)$ and the corresponding first differences $\Delta T(B)= T_{N,M}(B)-T_{N-1,M}(B)$ for $1\le N\le7$ and $M=7$ assuming a perfect lens potential without dip and QPCs perfectly alligned along the lens` principle axis. The focus point is varied from 
(a,d) behind the detector at $\vl=-0.47$\,V ($\Phi_\text L\simeq0.25\ef$), 
(b,e) on the detector at $\vl=-0.64$\,V ($\Phi_\text L\simeq0.50\ef$) and 
(c,f) between emitter and detector at $\vl=-0.85$\,V ($\Phi_\text L\simeq0.75\ef$).
}
\label{fig:TB-different-VL-ideal}
\end{figure*}

\section{Lateral mode structure near focal point}

Figures 3(a) and\ 3(b) of the main article display the measured vs.\ calculated transmission through both QPCs in series with $N=M=7$ as a function of $B$ and the lens gate voltage \vl. Modulations as a function of $B$ (vertical cuts, constant \vl) can be interpreted in terms of the lateral mode structure in the current profile emitted from a QPC. For $\vl=0$ we find strong modulations as a function of $B$, see also Fig.\ 2(a) of the main article. For $\vl=-0.64\,$V the emitted current is focused on the detector QPC and the modulation in $B$ almost completely disappears. As we tune \vl\ from $0$ via the focus condition to $\vl<-0.64\,$V the focal point moves from behind the detector onto the detector and finally in front of the detector. Hence, for $\vl<-0.64\,$V we see again a modulation in $B$.

For clarity we have also calculated this transition for the case of $1\le N\le7$ and $M=7$, for the same conditions as for Figs.\ 2(b) and 2(c) of the main article. In \fig{fig:TB-different-VL-ideal}{} we display $T(B)$ for three values of \vl, namely $\vl>-0.64\,$V in panel (a), $\vl=-0.64\,$V in (b)  $\vl<-0.64\,$V in (c) and then the corresponding first differences $\Delta T = T_{N,M}-T_{N-1,M}$ in the line below. Here we assumed a perfect lens without dip and perfectly aligned QPCs. Clearly the mode structure is almost gone under focusing condition for $\vl=-0.64\,$V. The qualitative differences between the modulation in $T(B)$ on the two sides of the focal point, i.e., for  $\vl>-0.64\,$V vs.\ $\vl<-0.64\,$V are a consequence of the lateral coherence of the ballistic electron beam. This interference effect is sensitive to the detailed electrostatic potential landscape of the lens which, in turn, depends on \vl.

\begin{figure*}[htb]
\includegraphics[width=2.0\columnwidth]{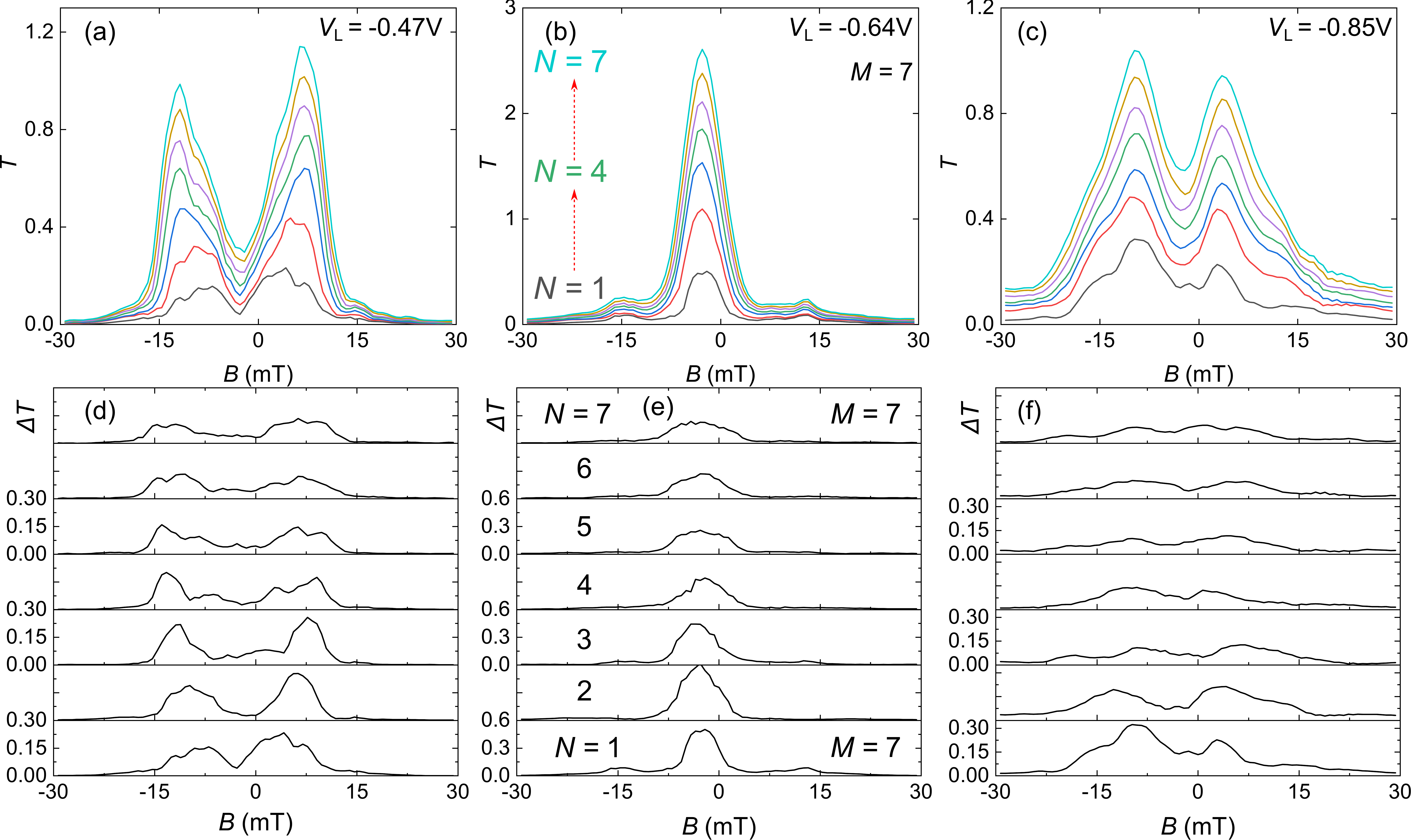}
\caption{Calculations at same conditions as in Fig.\ \ref{fig:TB-different-VL-ideal} beside that we include the electrostatic potential dip at the lens waist and the slight lateral misalignment of the QPC positions discussed   with the focus point being 
(a,d) behind the detector at $\vl=-0.47$\,V ($\Phi_\text L\simeq0.25\ef$),
(b,e) on the detector at $\vl=-0.64$\,V ($\Phi_\text L\simeq0.50\ef$) and 
(c,f) between emitter and detector at $\vl=-0.85$\,V ($\Phi_\text L\simeq0.75\ef$).
}
\label{fig:TB-different-VL-real}
\end{figure*}

In  \fig{fig:TB-different-VL-real}{}
we present the same calculations, but this time including our geometric imperfections, i.e., the electrostatic potential dip at the lens waist and the slight lateral shifts of the QPCs discussed in all detail below. As a result, the symmetry in $B$ is broken and details of $T(B)$ are modified. However, we still observe the same qualitative focusing effect at $\vl=-0.64\,$V.

\section{Measurement conditions and sample properties}\label{sec:conditions}

\subsection{Experimental setup}

The presented experimental data are all direct current (dc) measurements using the current amplifiers model 1211 of DL Instruments. As voltage sources for the gates and the source-drain voltage we used the model 7651 of Yokogawa. 
All measurements presented in the main article or the appendix were performed in a helium-3 evaporation cryostat and, if not stated otherwise, at a temperature of $T_\text B\simeq250\,$mK. For the presented data we applied a source-drain voltage of $V=-1\,$mV across the emitter QPC unless stated otherwise. Control measurements with $V=-0.1\,$mV at $T_\text B\simeq250\,$mK result in almost identical curves. Measurements with the temperature raised to 4\,K or more cause a broadening of the steps between conductance plateaus of the QPC pinch-off curves and, related,  a broadening of the oscillations in $I(B,\vl)$. 

We parameterize our experimental curves in terms of the mode numbers $N$ of QPC$_1$ and $M$ of QPC$_2$, which define the quantized conductance plateaus via $N,M=G_{1,2}/G_\text Q$ and $G_\text Q=2e^2/h$. If not stated otherwise, we adjust the applied gate voltages such that the QPCs are tuned close to the centers of the respective conductance plateaus. The measurement in Section~\ref{sec:measurement_QPCeigenmodes} demonstrates that this approach is viable.

\subsection{Inhomogeneous lens potential}\label{sec:lens_dip}

Our lens shows abberation. The main reason is a dip of the electrostatic potential located below the center of the lens gate. The potential dip is caused by strain below the lens gate which built up during cooldown due to the different thermal expansion coefficients of the metal gate and the (Al,Ga)As wafer. For the given lens geometry and its alignment in $\left<110\right>$-crystal direction the built-up strain generates a constant electric field via the piezoelectric effect. It gives rise to a symmetric potential dip along the lens (in the $y$-direction) which reflects the lens geometry. Comparison with literature confirms that the dip can lower the local electrostatic potential by a few meV \cite{Asbeck1984,Tanaka1997}. Experimentally we find a dip depth of $\Delta_\text{dip}\simeq4.2\,$meV, see the end of Section~\ref{sec:calibration}. This potential modulation is a general feature of on-chip electrostatic lenses on a piezoelectric material and should be taken into account for future lens designs. 

\subsection{Misalignment of the nanostructures}\label{sec:nanostructure-shifts}

\begin{figure}
\includegraphics[width=0.8\columnwidth]{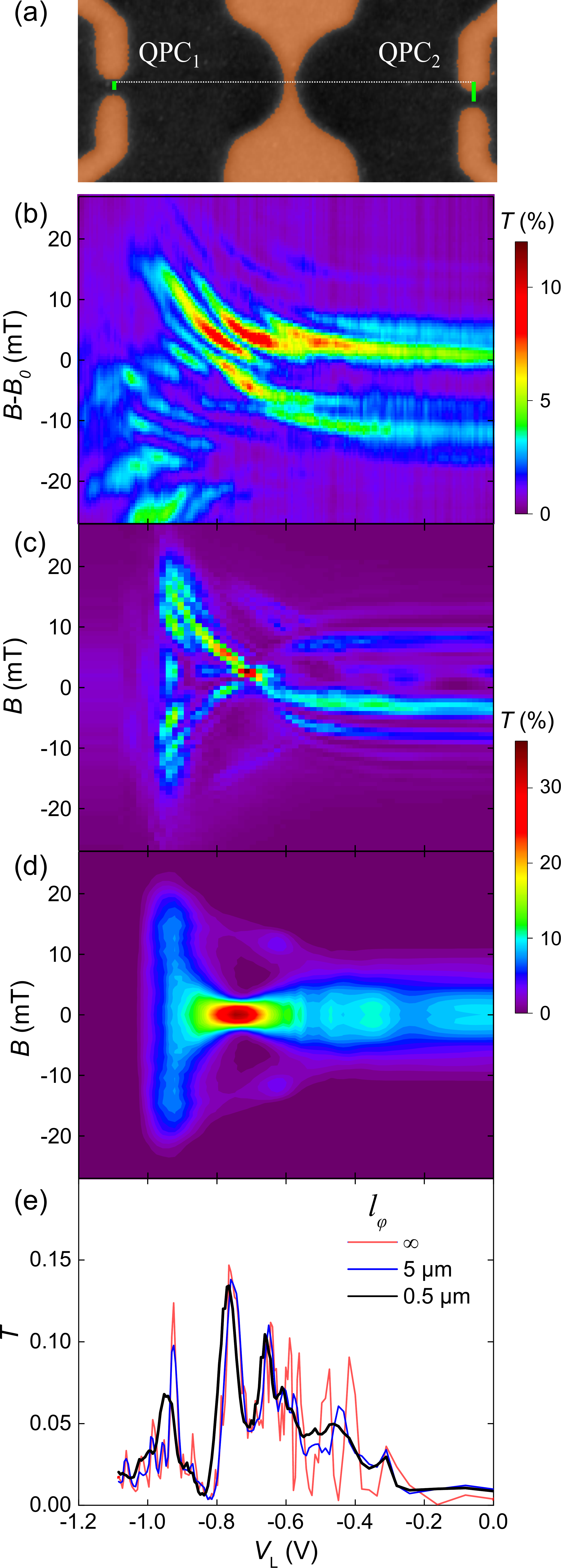}
\caption{
(a) AFM image of the sample surface. Light color corresponds to the metal gates. Green bars depict the lateral shifts of the QPCs from the waist of the lens (dashed horizontal line) by $80$\,nm and $260$\,nm, respectively. 
(b) Measured transmission $T(B,\vl)$ through both QPCs in series at $N=M=1$ at a source drain voltage $V=-1$mV. 
(c) Calculated  $T(B,\vl)$ for $N=M=1$ for the lateral misalignments measured in (a) including a electrostatic potential dip at the lens waist as discussed in section \ref{sec:lens_dip} .
(d) Calculated  $T(B,\vl)$ for $N=M=1$ for a perfectly symmetric sample without electrostatic potential dip at the lens waist.
(e) Calculated $T(B=0,\vl)$ cuts for $N=M=1$ as in (b) but with varying dephasing length $l_\phi$. 
} 
\label{fig:nanostructure-shifts}
\end{figure}

In \fig{fig:nanostructure-shifts}{a}
we present an atomic force microscope (AFM) image of the sample. It reveals that both QPCs are shifted [downwards in \fig{fig:nanostructure-shifts}a] with respect to the lens waist, QPC$_1$ by $\simeq80$\,nm and QPC$_2$ by $\simeq260$\,nm (green bars). For the calculations we used slightly smaller shifts for a slightly better agreement with theory. As discussed in the main paper, a lateral misalignment of the two QPCs in respect to the lens breaks the symmetry and together with the electrostatic potential dip reduces the mode-to-mode coupling and the ballistic transmission through the two QPCs in series. The effect is strongest at small values of $N$ and $M$ because the lower modes emit narrower electron beams. In \fig{fig:nanostructure-shifts}{b} we present the measured serial transmission $T(B,\vl)$ for $N=M=1$. The transmission is small and asymmetric with respect to $B-B_0=0$. 
An according numerical calculation in panel (c) qualitatively reproduces this behavior. For clarity, in panel (d) we also present the model prediction for a hypothetical sample with perfectly positioned QPCs and a lens without electrostatic potential dip. 

\subsection{Influence of the quantum point contact eigenmodes} \label{sec:measurement_QPCeigenmodes}

\begin{figure}[htb]
\includegraphics[width=0.75\columnwidth]{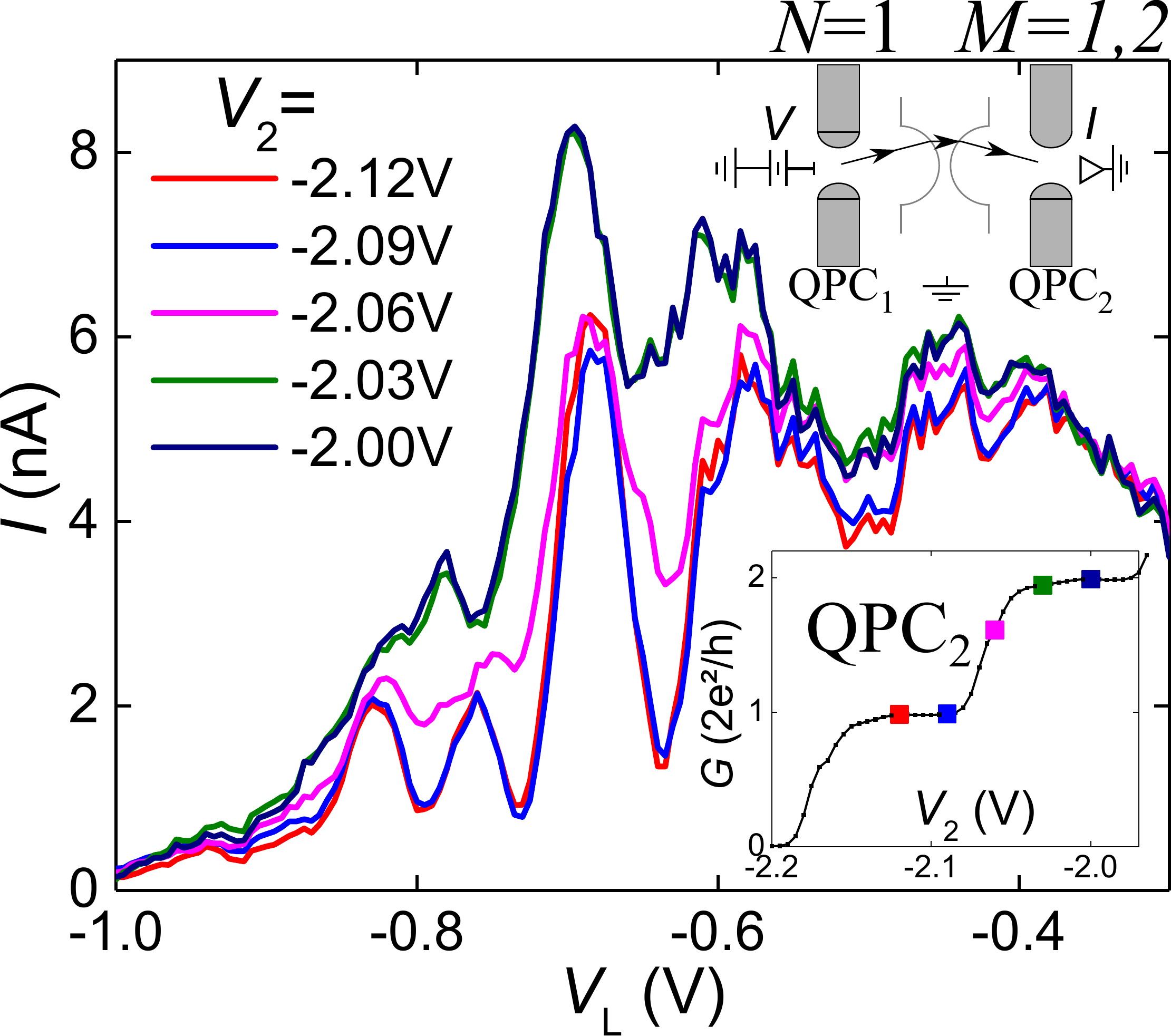}
\caption{Measured detector current $I(\vl)$ at $B=B_0$ for the emitter (QPC$_1$) tuned to the center of the first plateau ($N=1$) at source-drain voltage $V=-1$\,mV while the detector (QPC$_2$) is tuned to various values between $M=1$ and $M=2$ by varying $V_2$. The detector configurations are indicated in the lower inset.
\label{fig:Lens-N=1_M=1,2-transition}}
\end{figure}
The lower inset of \fig{fig:Lens-N=1_M=1,2-transition}{} plots the detector pinch-off curve for $M\le2$. In the main figure we present the detector current, i.e., the current through both QPCs in series as a function of the lens gate voltage $V_\text L$ at $B=0$. The emitter QPC$_1$ is tuned to its first conductance plateau ($N=1$) while the detector is tuned to various points between $1\le M\le2$ as indicated in the inset. Traces $I(\vl)$ measured with the detector on one and the same plateau are almost identical even in small details. In contrast, we observe large variations between traces with different detector conductance. This result demonstrates that the current profile between the QPCs is directly related with the occupied eigenmodes of the QPCs. The latter stay unchanged as long as both QPCs remain on their respective plateaus. However, the mode structure and current profile rapidly change as soon as the conductance of one of the QPCs is changed.

\subsection{Influence of coherent electron dynamics}
\label{sec:origin_osc_TVl}

The traces $I(V_\text L)$ in \fig{fig:Lens-N=1_M=1,2-transition}{} display strong oscillations, which depend on the mode occupation of the QPCs, i.e., $N$ and $M$. They are stronger for smaller $N, M$, lower temperature and smaller source drain voltage $V$, indicating an origin in terms of coherent standing waves (i) between the two QPCs or (ii) between emitter or detector QPC and the lens. Furthermore, (iii) the presence of the electrostatic potential dip at the center of the lens causes diffraction of the electron beam resulting in a modulated transmission profile $T(V_\text{L}=0, B)$ as seen both in experiment, Figure \ref{fig:nanostructure-shifts} panel (b), and in the simulation Figure \ref{fig:nanostructure-shifts} panel (c) as bent stripes of local transmission maxima. Due to the asymmetry of the sample, this diffraction results in oscillations of the transmission along horizontal cuts of the transmission profile, e.g., for $B=0$. 

As the interference pattern strongly depends even on slight disorder and the precise realization of the electrostatic potential, a one-to-one comparison with a numerical simulation is difficult. In order to qualitatively explain the oscillations in $I(V_\text L)$ in \fig{fig:Lens-N=1_M=1,2-transition}{}, we compare simulated $T(V_\text{L}, B=0)$ cuts with varying dephasing length $l_\phi = 0.5\,\mu$m, $l_\phi = 5\,\mu$m and $l_\phi = \infty$ shown in \ref{fig:nanostructure-shifts} (e). The fast oscillations for small negative lens voltages $\vl > - 0.6$V are consistent with an interpretation in terms of standing waves between the two QPCs as they are strong for a dephasing length much larger than the distance between the QPCs $l_\phi = \infty$ while they are averaged out for a dephasing length $l_\phi\le 5\,\mu$m of the order of the distance between the QPCs. The oscillations at large negative lens voltages $\vl < - 0.6$V can originate both from (ii) coherent standing waves between the QPC and the lens and (iii) the lens voltage tuning the diffraction pattern through the $B=0$ cut. The former exist only if the lens voltage $\vl$ is negative enough such that electrons approaching a sufficiently smooth electrostatic potential $\Phi (\vl)$ are reflected if their forward momentum is smaller than $\sqrt{2 m^* ( E_\text{F}^0 - \Phi_\text{L})}$. 
As the experiment was conducted at $T=250\,$mK with a source drain voltage of $-1\,$mV corresponding to a dephasing length of $0.5\,\mu$m, oscillating current contributions due to standing waves between lens and the QPCs are averaged out. Thus, the simulation supports the conclusion that the large oscillations in the experimental data at $\vl < -0.6\,$V of Fig. \ref{fig:Lens-N=1_M=1,2-transition} are caused by (iii) the tuning of the diffraction pattern through the $B=0$ cut. This conclusion is in agreement with the measured and simulated $T(\vl,B)$ profile for emitter and detector plateaus $N=M=1$ in Fig. \ref{fig:nanostructure-shifts} (b) and (c), respectively. Note, that the simulated $T(\vl,B)$ profile in a symmetric geometry without electrostatic potential dip at lens waist does not exhibit these oscillations.

\subsection{Current density modulation due to the piezoelectric potential dip}

To further demonstrate the impact of the piezoelectric potential dip at the lens waist, we calculate the current density emitted from a hard-wall QPC for $N=1$ and $N=7$ in the presence of the electrostatic potential dip but an otherwise flat potential. The results are shown in Fig.~\ref{fig:beamprofiles_dip}.
{\begin{figure}
\includegraphics[width=0.8\columnwidth]{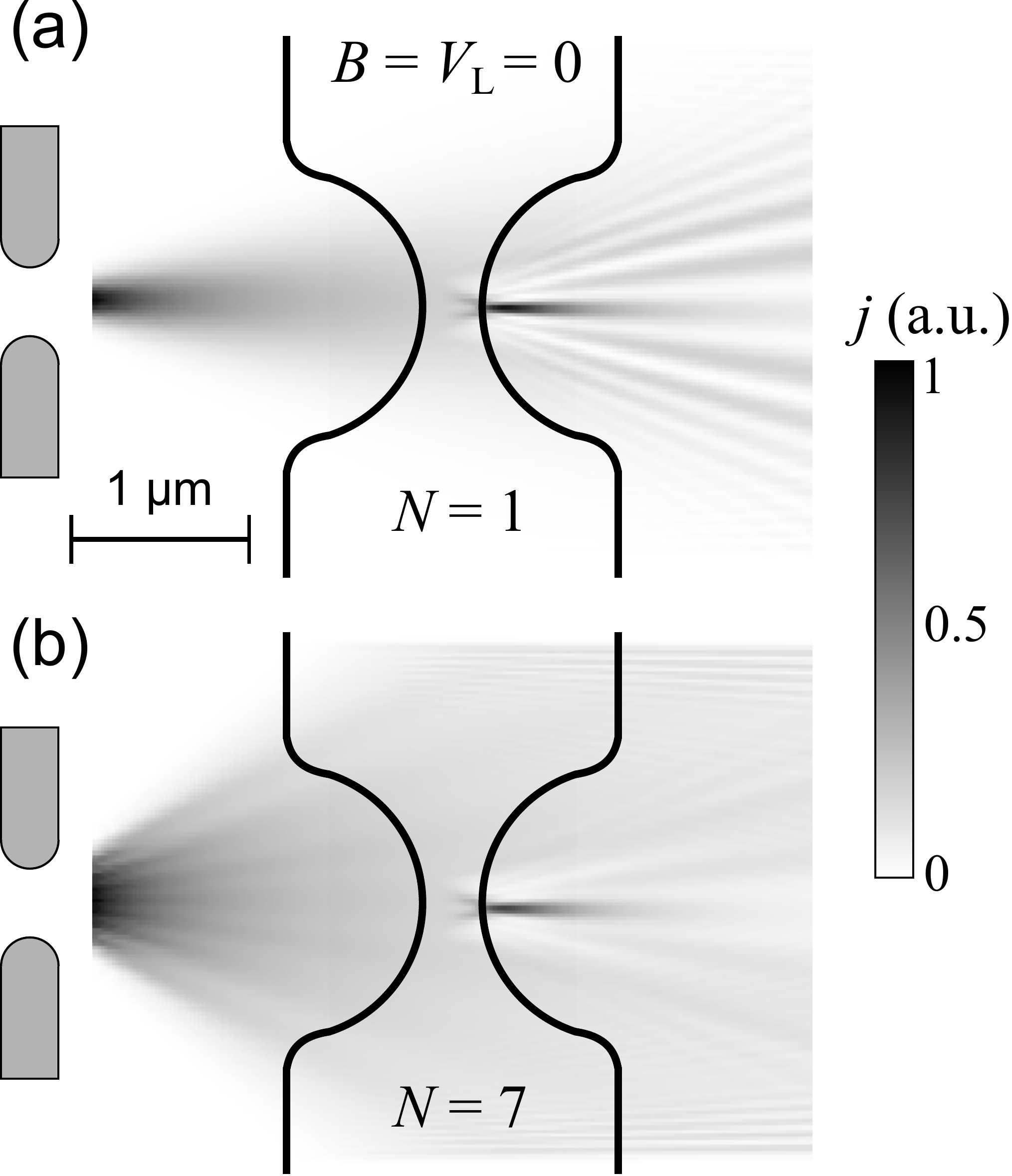}
\caption{Current density emitted from the hard-wall QPC (a) at $N=1$ and (b) at $N=7$ accounting for the piezoelectric potential dip at the lens waist but an otherwise flat potential. We sketch the edge of the lens gate for comparison.} 
\label{fig:beamprofiles_dip}
\end{figure}}

For $N=1$ the calculated current density emitted from the QPC is a gaussian beam. Scattering off the dip gives rise to a diffraction pattern in the current density. 
Signatures of this diffraction pattern are visible both in the experimental and theoretical current profiles $T(\vl, B)$, c.f.\ Fig.\ \ref{fig:nanostructure-shifts}(b) and (a), as oscillating current maxima for fixed $\vl$, in particular for lens gate voltages above the focal point. 
Comparison to the calculated current density and current profiles $T(\vl, B)$ in the absence of the potential dip, c.f.\ Fig.\ \ref{fig:beamprofiles}(a) and Fig. \ref{fig:nanostructure-shifts}(d), respectively, shows that the observed oscillations discussed here are indeed a signature of diffraction from the potential dip.

For $N=7$, the current density emitted from the QPC is a superposition of all seven QPC modes. As electrons emitted in different modes are uncorrelated, the diffraction pattern due to scattering off the dip averages out details of the individual modes and is much less pronounced compared to $N=1$. 
The main consequence of the presence of the dip is a reduction of the current along the principal axis (besides for a small distance, where the current density at the principal axis is enhanced). In particular, the maxima at $B\simeq\pm15\,$mT of our measured $I(B)$ data at $\vl = 0$, c.f.\ Fig.\ 2(a) of the main paper or Fig.\ \ref{fig:I(B)_det-M=7_em-N=1,...,7}, can be explained with scattering off the electrostatic potential dip. 
In comparison, the calculations without potential dip, c.f.\ Fig.\ \ref{fig:beamprofiles}(b) and \fig{fig:QM_vs_Classical}{}, respectively, does not show pronounced maxima at $B\simeq\pm15$\,mT but instead displays a current maximum at $B=0$.

A direct comparison between our experimental current profiles $I(B)$ at $\vl = 0$, c.f.\ Fig.\ 2(a) in the main paper, and the calculations without the electrostatic potential dip, c.f.\ Fig.\ \ref{fig:QM_vs_Classical}, indicates that our experimental results are strongly influenced by the presence of the dip.

Finally, we wish to point out that the observed distribution of minima and maxima in the current differences $\Delta T$, shown in Figs.\ 2(b) and 2(c) of the main paper, is a clear signature of the mode structure of the QPCs. In our calculations we cannot model the observed pattern of oscillations for all $1\le N\le7$ based on scattering off the electrostatic dip. This argument still holds, even if we include disorder scattering assuming identical disorder potentials for all $N$.

\section{Lens calibration}\label{sec:calibration}

The variations between the dispersion relations of light with the momentum $\propto1/c$ and that of massive particles results in different refraction laws. The momentum of the relevant electrons is in our case proportional to its Fermi velocity \vf. In optics, the refractive index of a medium is defined as the ratio of the vacuum light speed $c$ to the phase velocity $c_\text m$ of light in the medium, $n_\text m=c/c_\text m$. In analogy, we define the relative refractive index of our electrostatic lens as $\nr=\vf^0/\vfl$, where $\vf^0$ is the Fermi velocity of electrons in the bulk 2DES away from the lens and \vfl\ its counterpart below the lens gate. 

The focusing properties of a lens can be calculated by requiring a smooth transition between regions of different refractive index, i.e.\ that the component of the momentum in the plane of the interface of changing refractive index remains constant. Snell's law for light then reads
\begin{equation}
\text{for photons:\quad}n_\text m=\frac c{c_\text m}=\frac{\sin\delta_1}{\sin\delta_2}\,,
\end{equation}
where $\delta_1$ is the angles of incidence (in vacuum) and $\delta_2$ the angle of refraction (in the lens). In contrast for electron optics we find a different refraction law \cite{Spector1990,Sivan1990}
\begin{equation}\label{eq:snell}
\text{for electrons:\quad}\nr=\frac{\vf^0}{\vfl}=\sqrt{\frac{\ef^0}{\efl}}=\frac{\sin\delta_2}{\sin\delta_1}\,,
\end{equation}
where we introduced the kinetic energies $\ef^0$ and \efl\ in the bulk 2DES away from the lens and below the lens gate. As a consequence, to achieve focusing for $\nr>1$, an optical lens must be convex while an electrostatic lens must be concave.

Quantitative predictions of the focusing properties of our lens require a calibration of the electrostatic potential $\Phi_\text L=\ef^0-\efl$ induced by the lens gate as a function of the voltage \vl\ applied to it. Below, we compare two different methods to experimentally calibrate the lens. Comparison between the two methods reveals the dip of the lens potential discussed in Sec.\ \ref{sec:lens_dip} above.

\subsection{Landau-level lens calibration}
\label{sec:LL_calibration_lens}

A viable method to calibrate the height of a barrier is to measure the reflection of quantum-Hall edge-channels in a strong perpendicular magnetic field $B$. The edge channels are a consequence of the orbital quantization of the density of states of the 2DES in Landau levels (LLs) with quantum number $l=1, 2, 3, \dots$. Taking into account the Zeeman splitting each LL results in two spin-resolved energy levels quantized at   
\begin{equation}\label{eq:LL-energies}
E_\pm(l)=\hbar\omegac\left(l-\frac12\right)\pm\frac12g\mu_\text BB\,,
\end{equation}
where $\omegac=eB/m^{*}$ is the cyclotron frequency, $e$ is the elementary charge, $m^\star=0.067m_0$ is the effective electron mass in GaAs with the free electron mass $m_0$, $g$ the Landee g-factor with $g\simeq-0.36$ in GaAs and $\mu_\text B$ the Bohr magneton. We indicate the corresponding spin polarized edge channels by their filling factor $\nu=1,2,3,\dots$ defined as $\nu=2 \ef^0 /\hbar\omegac$. For instance at $\nu=2.25$ the lowest two edge channels, corresponding to the lowest LL, are completely filled and one quarter of the states of  (spin-up polarized) third level are also occupied.

\begin{figure}[htb]
\includegraphics[width=1\columnwidth]{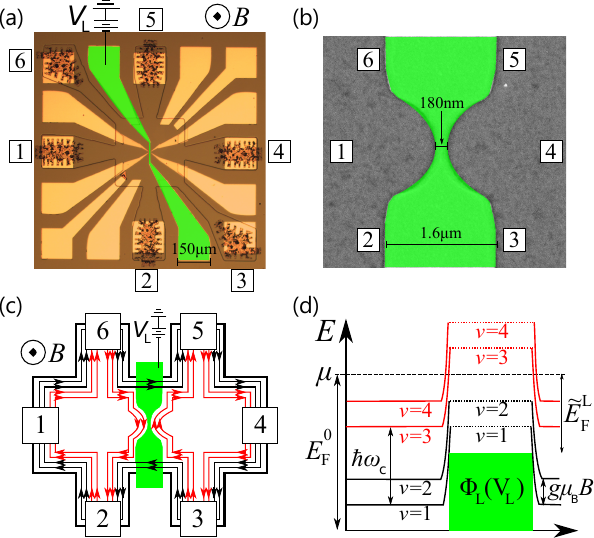}
\caption{(a) Photograph of the sample. Six ohmic contacts (spotty meander structures) at the periphery are numbered consecutively. The 2DES containing mesa is surrounded by a thin black line. Golden metal gates are used to define the two QPCs while the lens gate is colored in green.
(b) Scanning electron micrograph of the central lens gate (green) on the GaAs surface (gray). Locations of ohmic contacts are indicated by the numbers.
(c) Sketch of current carrying edge channels connecting the ohmic contacts 1--6 in the quantized Hall regime at integer filling factor $\nul=2$ beneath the lens gate and $\nu=4$ elsewhere.
(d) Illustration of the reflection of edge channels. Edge channels corresponding to the red quantized energy levels (with $\nu=3, 4$) are reflected at the lens barrier if its height caused by applying a negative voltage $\vl>0$ to the lens gate fulfills $\Phi_\text L\ge\hbar\omegac$.
}
\label{fig:LL-cal-setup}
\end{figure}

In our calibration measurements we apply a constant $B$ such that the bulk 2DES (away from the gates) has an integer filling factor. By sweeping \vl\ we then gradually increase the lens barrier, hence decrease \efl\ and the filling factor beneath the lens gate $\nu_\text l$. Thereby, we detect the successive reflection of the individual edge channels at the barrier in terms of the corresponding resistance changes. In \fig{fig:LL-cal-setup}{a} 
we present an overview photograph of the wafer surface including bond pads and ohmic contacts in the periphery. The position of the ohmic contacts, labeled by numbers 1--6, in respect to the lens gate are also indicated in panels (b) and (c), which show an SEM image of the central part of the lens gate and a sketch of the quantum Hall measurement set-up, respectively. In panel (c) we assume the filling factor of $\nu=4$ in the bulk but $\nu_\text l=2$ below the lens gate, such that the third and the fourth edge channel (red) are reflected off the lens. As illustrated in \fig{fig:LL-cal-setup}d, this reflection occurs for $\efltilde<\nu_\text L/2\,\hbar\omegac=\hbar\omegac$. Here \efltilde\ denotes the maximum of $\efl(y)$ taking into account a possible dip of $\Phi_\text L(y)$ in the center of the lens. In our sketch \efltilde\ is centered in the gap between two LLs, such that below the lens gate [precisely at the minimum of $\Phi_\text L(y)$] all states corresponding to filling factors $\nu=1$ and 2 are occupied while states at higher energies, e.g.\ for $\nu=3$ and 4, are empty. Based on the Landauer-B\"uttiker approach \cite{Buettiker1988,Haug1988} we expect to find the longitudinal resistance measured between contacts 3 and 2 across the lens
\begin{figure*}[t]
\includegraphics[width=1.9\columnwidth]{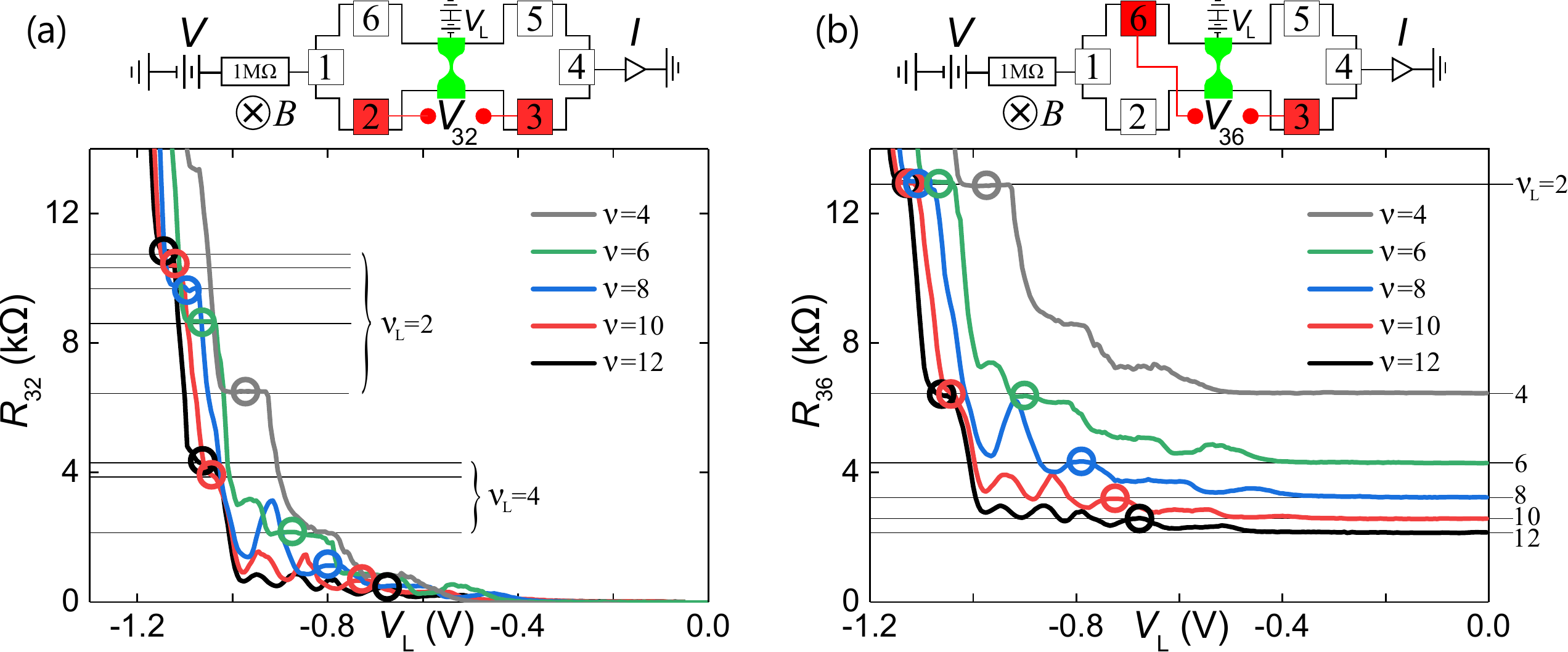}
\caption{Longitudinal resistance $R_{32}(\vl)$ defined in \eq{eq:Rlong} in panel (a) and transverse resistance $R_{36}(\vl)$ defined in \eq{eq:Rtrans} in (b). Each curve has been measured at constant $B$ and fixed bulk filling factor $\nu$ with $V=-400\,$mV applied at a $1$M$\Omega$ series resistor causing an approximately constant current of $I\simeq400$\,nA flowing from contact 1 into the grounded contact 4. Horizontal lines indicate plateaus expected at filling factors $(\nu,\nul)$ according to \eq{eq:Rlong} in (a) and \eq{eq:Rtrans} in (b) [not all expected plateaus are shown]. Our calibration points in the centers of the open circles are at identical values of \vl\ in (a) and (b).
Top sketches: Four-terminal circuits of our Landau-level calibration measurements.
}
\label{fig:Calibration_Resistance}
\end{figure*}
\begin{equation}\label{eq:Rlong}
R_{32}=\frac{V_3-V_2}{I_{41}}=\frac he^2 \left(\frac 1\nu_\text L-\frac 1\nu\right)\,,
\end{equation}
while applying a constant current between contacts 1 and 4. Likewise, between contacts 3 and 6 we expect to measure the transversal resistance across the lens gate
\begin{equation}\label{eq:Rtrans}
R_{36}=\frac{V_3-V_6}{I_{41}}=\frac he^2 \frac 1\nu_\text L\,.
\end{equation}
For simplicity and without a considerable loss of accuracy we restrict the analysis of our calibration experiment to even filling factors $\nu$ and $\nu_\text L$ (and disregard features allocated to odd  $\nu_\text L$). In \fig{fig:Calibration_Resistance}a we present the results of four terminal resistance measurements for the setups described by \eq{eq:Rtrans} and in panel (a) and according to \eq{eq:Rlong} in panel (b). Each curve has been measured at constant $B$ and bulk filling factors $\nu=4, 6, 8, 10, 12$ while decreasing the filling factor beneath the lens gate by sweeping \vl\ from 0 to the pinch-off point at $-1.2$\,V. 

\begin{figure}[h!]
\includegraphics[width=0.9\columnwidth]{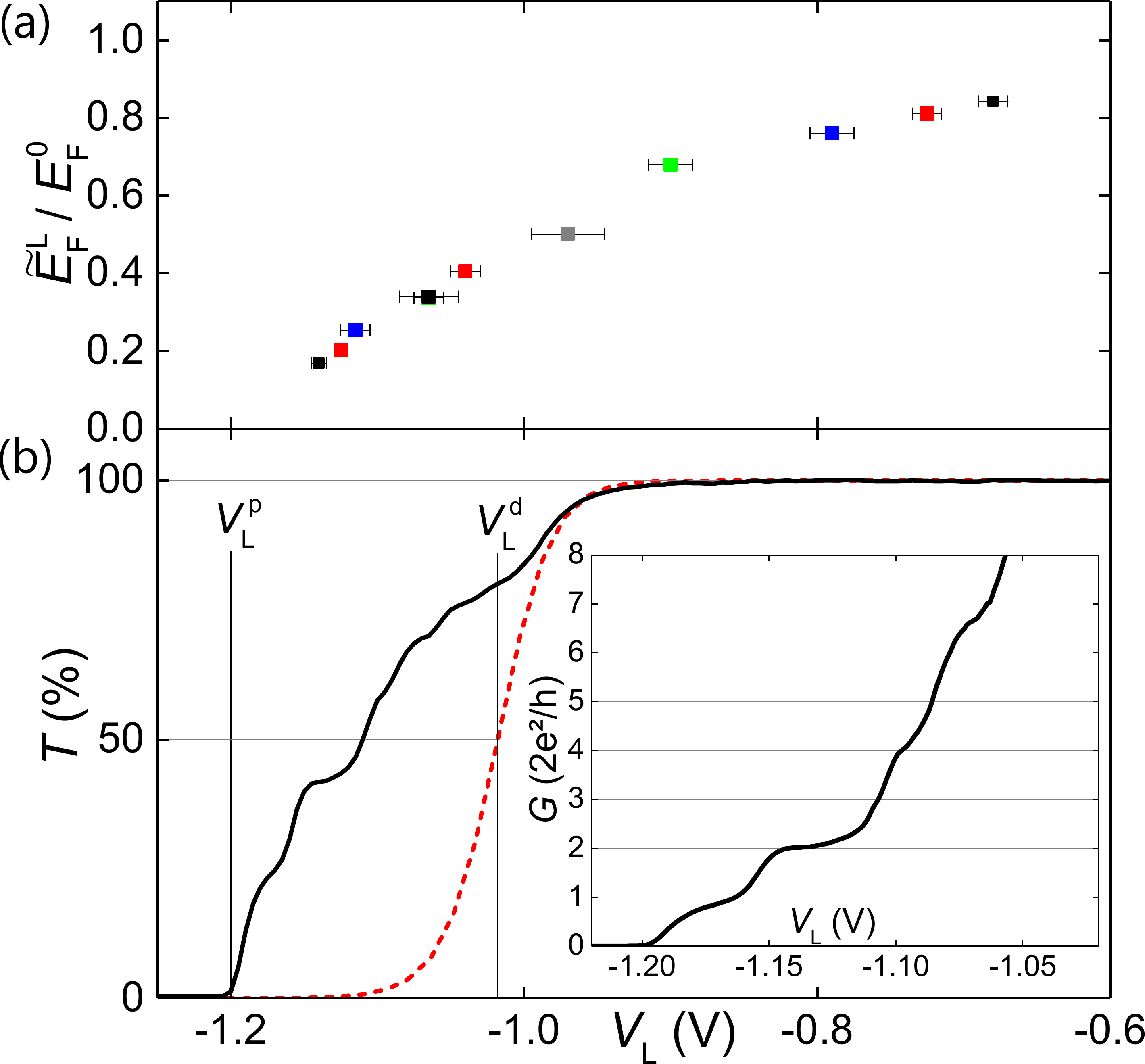}
\caption{(a) Calibration by edge channel reflection. Each point corresponds to a circle in Fig. \ref{fig:Calibration_Resistance}{} (same colors). Error bars correspond to the plateau widths in Fig. \ref{fig:Calibration_Resistance}{}.
(b) Solid black line: measured lens pinch-off curve at $B=0$ and no additional QPCs defined ($V_1=V_2=0$). The pinch-off voltage is $V^\text{p}_\text{L}\simeq-1.2$\,V. Dashed red line: calculated transmission through a (one-dimensional $250$\,nm wide parabolic) barrier fitted to the depletion onset of the lens’ pinch-off curve. It provides a rough estimate of the depletion voltage $V_\text{L}^\text{d}\simeq-1.02$\,V where the 2DES is depleted beneath wide gates, i.e.\ outside of the central dip. 
Inset: For $\vl\lesssim V_\text{L}^\text{d}$ the measured pinch-off resembles quantized conductance steps pointing to a one-dimensional conducting channel through the central dip.
} 
\label{fig:LL-Calibration+pinchoff-determination}
\end{figure}

Instead of the expected monotonous steps between resistance plateaus we often find local resistance maxima. This phenomenon is related with an additional tunnel current between reflected edge channels via localized states below the thin barrier and has also been observed in Ref.\ \onlinecite{Haug1988}. Where our measurements are close enough to the expected resistance plateaus (at the resistance values indicated by horizontal lines) we choose the center of these plateaus as calibration points $\vl(\nul)$ with
\begin{equation}\label{eq:LL-calibration}
\frac{\efltilde}{\ef^0}=\frac{\nu_\text L}{\nu}\,,
\end{equation}
see \fig{fig:LL-Calibration+pinchoff-determination}{a} for the final calibration result. As expected, the calibration points obtained in this way turn out to be almost identical for the two alternative methods presented in \fig{fig:Calibration_Resistance}{}.
\begin{figure*}[tb]
\includegraphics[width=2.0\columnwidth]{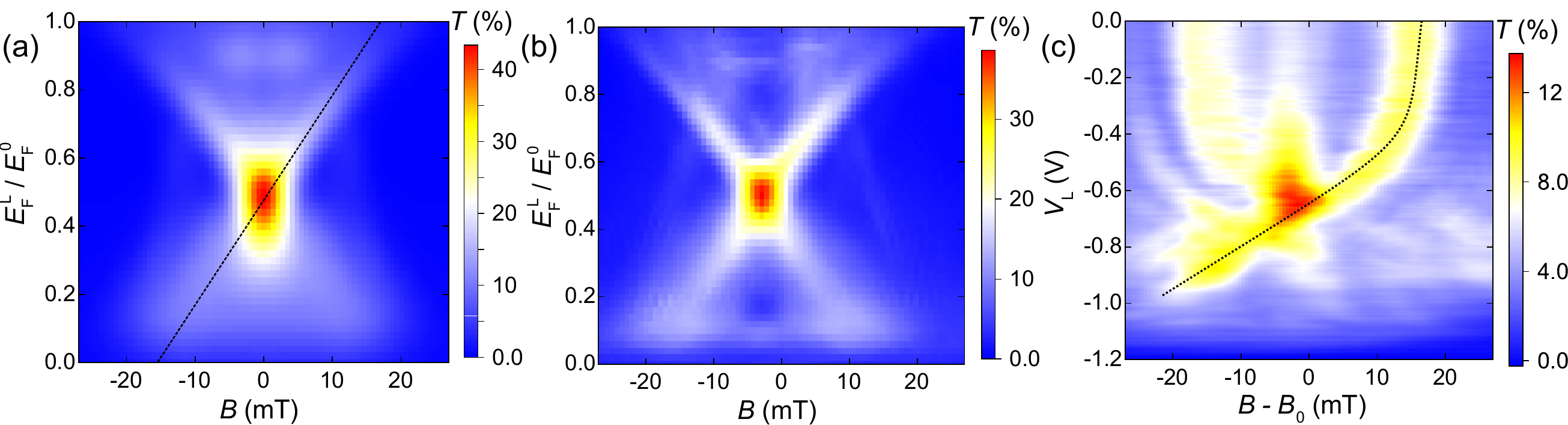}
\caption{(a) Calculated serial transmission $T(\efl,B)$ through two QPCs tuned to the 7th conductance plateau for a symmetric sample and a flat lens potential without potential dip. A dashed black line traces one of the main current maxima.
(b) $T(\efl,B)$ including the slight lateral shifts of the QPCs and the lens (cf. section \ref{sec:nanostructure-shifts}) and the  potential dip at the lens waist. (c) Measured transmission $T(\vl,B)$. A dashed black line traces one of the main current maxima. A point-by-point scaling of the two lines in (a) and (c) results in the calibration $\efl(\vl)$.} 
\label{fig:selfconsistent-Calibration}
\end{figure*} 
The dependence $\efltilde(\vl)$ is a fingerprint of our heterostructure. The non-linear relation reveals a corresponding lens voltage dependent capacitance between the lens gate and the 2DES beneath. We conjecture that this behavior is related with a high-resistance conducting layer, a delta doped layer located between the wafer surface and the 2DES. Because the carrier density (and mobility) in this layer, too, depends on the gate voltage it causes a \vl-dependent screening between lens gate and 2DES.

In \fig{fig:LL-Calibration+pinchoff-determination}{b} we show (as a solid line) the pinch-off curve $G(\vl)$ of the lens gate measured at $B=0$. Related with the concave shape of the lens gate it develops a few quantized conductance steps at integer multiples of $G_\text Q$, emphasized in the inset. They indicate a potential dip at the center of the lens forming a weakly confined QPC. Such a dip can be explained in terms of the piezoelectric effect of GaAs which gives rise to a sizable inhomogeneous build-in electric field below the lens gate. The piezo-field is caused by strain in response to stress at the metal-semiconductor interface built up during cool-down because of the different thermal expansion coefficients of the materials. The detailed build-in field depends on the geometry and orientation (as the piezo tensor is anisotropic) of the lens but can alter the potential $\Phi_\text L\text (x,y)$ locally by up to several meV \cite{Asbeck1984,Tanaka1997}. 
Our LL calibration measures the absolute minimum of the barrier height, reduced at the dip located at the center of the lens. However, the focusing properties of a lens are naturally predominantly determined by its curvature and potential further away from its very center.

\subsection{Self-consistent calibration}
\label{sec:self-consistent_calibration}

\begin{figure}[htb]
\includegraphics[width=0.75\columnwidth]{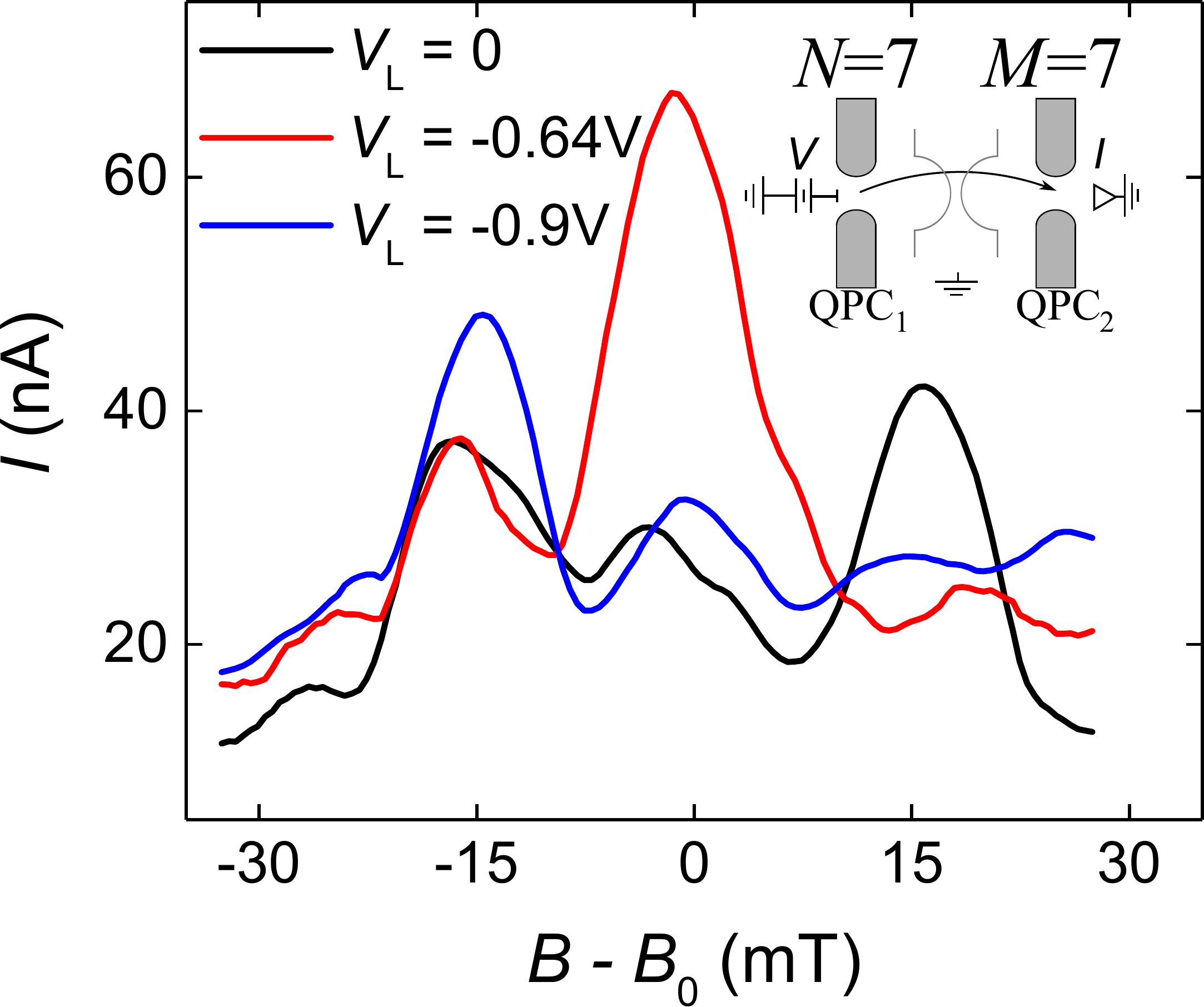}
\caption{Measured detector current $I(B)$ for $N=M=7$ for three different values of \vl. The inset illustrates the three-terminal circuit.
\label{fig:I(B)-VL-cuts}}
\end{figure}

Our second calibration method is more suited to determine the lens potential away from its center. It relies on a self-consistent comparison of the measured focusing properties of the lens with our theoretical predictions. In \fig{fig:selfconsistent-Calibration}{} we present the transmission $T$ through both QPCs in series and tuned to their 7th conductance plateaus ($N=M=7$). In panel (c) we re-plot the measured data $T(B,\vl)$ already presented in Fig.\ 3(a) of the main article. Tracking and comparing the measured versus predicted current maxima as function of $B$ and \vl\ versus \efl, cf.\ dashed lines, provides the seeked relation for $\efl(\vl)$. This tracking procedure relies on identifying the absolute current maximum as a function of $B$. To demonstrate the feasibility of this procedure we present in \fig{fig:I(B)-VL-cuts}{}
$I(B)$ for three different values of \vl, corresponding to three vertical cuts through Fig.\ 3(a) of the main article. The absolute current maximum moves from positive to negative values of $B$ as \vl\ is decreased. At $\vl=-0.64\,$V (red line) the current at $B-B_0=0$ is strongly enhanced, indicating electrostatic focusing. 

Note, that the tracked current maximum for $N=M=7$ corresponds to a trajectory between the QPCs which does not touch the very center of the lens. This way we try to avoid the influence of the dip at the center of the lens for the present calibration. Consequently, in our calculations we assume a homogeneous (flat) lens potential $\Phi_L(\vl)=\ef^0-\efl(\vl)$ independent of $y$ but disregard the dip in the center of the lens.

In addition to tracking the current maximum, we use the following prominent points to adjust the quantitative calibration result: (i) at the focus point measured at $\vl=-0.64$\,V our calculation predicts $\efl=0.49\ef^0$; (ii) we assume that the lens gate (away from its center, where it has a dip) has a negligible effect on the local potential at $\vl=0$: $\Phi_\text L(\vl=0)=0$; and (iii) as discussed above we estimate the pinch-off point for the lens away from the center dip to occur at $\vl\simeq-1.02$\,V which leads to $\Phi_\text L(\vl\simeq-1.02$\,V$)=\ef^0$.

\begin{figure}[tbh]
\includegraphics[width=0.9\columnwidth]{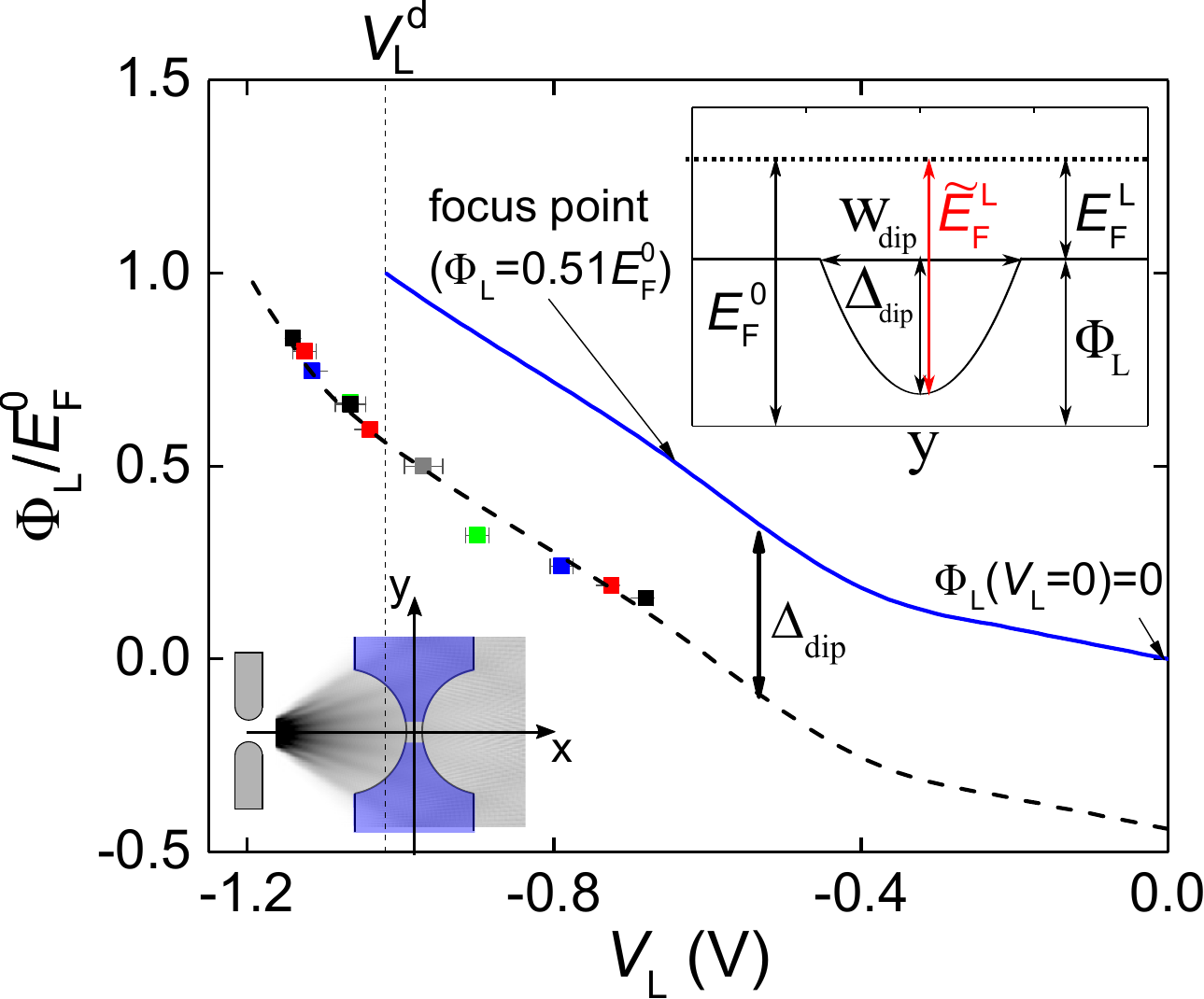}
\caption{Comparison between the two calibration methods, plotting $\Phi_\text L/\ef^0=1-1/\nr^2$. The blue solid line indicates the result of the self-consistent and the colored symbols that of the LL lens calibration. The dashed black line is parallel to the solid line but vertically shifted to fit the LL calibration points. The vertical shift corresponds to the dip depth $\Delta_\text{dip}$.
Top right inset: lateral lens potential with parabolic dip of width $W_\text{dip}$ and depth $\Delta_\text{dip}$ in its center. The lens potential $\Phi_\text L$, \efl\ and $\efltilde=\Delta_\text{dip}+\efl$ depend on \vl\ while the bulk Fermi energy $\ef^0$ is a material property. The corresponding dip curvature is $\omega_\text{dip}=1.06\times10^{12}$\,s$^{-1}$, its depth $\Delta_\text{dip}=4.2$\,meV and its width $W_\text{dip}=270$\,nm. 
Bottom left inset: calculated quantum mechanical current density emitted from QPC$_1$ with $N=7$. The outer flanks of the beam miss the dip of the lens potential and are unaffected by it. 
} 
\label{fig:selfconsistent-Calibration+dip}
\end{figure}

In \fig{fig:selfconsistent-Calibration+dip}{}
we present the obtained relation as a solid line in comparison with our LL lens calibration (symbols). 
Assuming in our calculations a flat lens potential $\Phi_L(\vl)=\ef^0-\efl(\vl)$ independent of $y$, we disregard the dip in the center of the lens. As such, the self-consistent calibration approximately averages the lens potential weighted by the actual lateral current distribution (in $y$-direction). For $N=M=7$ the lateral current distribution has two pronounced maxima away from the center of the lens, as visible, e.g., in the inset of \fig{fig:selfconsistent-Calibration+dip}{}. Underpinned by measured data, we argue that the current density is much smaller in the dip region, such that our second calibration mostly probes the lens potential away from its center.

\subsection{Determination of the inhomogeneous lens potential}

With our second calibration method we predominantly determined the lens properties away from its center, hence outside its potential dip, while with our first calibration method we measured precisely the minimum of the electrostatic lens potential. Hence, we interpret the difference between the two results as being the depth of the dip. In \fig{fig:selfconsistent-Calibration+dip}{} it is indicated as $\Delta_\text{dip}$. Our data are consistent with a constant $\Delta_\text{dip}(\vl)$, i.e., independent of \vl. This is expected if the electric field is generated by strain combined with the piezoelectric effect.

The pinch-off curve of the lens plotted in \fig{fig:LL-Calibration+pinchoff-determination}{b} provides additional information about the shape of the central potential dip. Using the LL lens calibration we determine from the step width between the two lowest quantized conductance plateaus the corresponding one-dimensional subband spacing: $\Delta E_{01}\simeq0.7$\,meV. For a rough estimation we assume a parabolic potential dip in $y$-direction, which is centered in an otherwise flat barrier:
\begin{align*}
\Phi_\text{dip}(y) &= \Phi_L(\vl)-\Delta_\text{dip}+\dfrac{1}{2}m^\star \omega_\text{dip}^2 y^2\, &\text{for }|y|<w_\text{dip}/2\\
\Phi_\text{dip}(y) &= \Phi_L(\vl)\, &\text{for }|y|\ge w_\text{dip}/2
\end{align*}
where $\omega_\text{dip}=\Delta E_{01}/\hbar$ is its curvature and $W_\text{dip}=2\hbar/\Delta E_{01}\sqrt{2\Delta_\text{dip}/m^\star}\simeq270$\,nm its width at $\Phi_\text{dip}(y)=\Phi_L(\vl)$, \fig{fig:selfconsistent-Calibration+dip}{}. The dip width is indicated by the central gray region of the lens gate in the inset of \fig{fig:selfconsistent-Calibration+dip}{}. It is small enough to corroborate our assumption that the dip has only little influence on our second calibration method.

In  \fig{fig:Imp-vs-Imp+Dip}{}
{\begin{figure}[tbh]
\includegraphics[width=0.9\columnwidth]{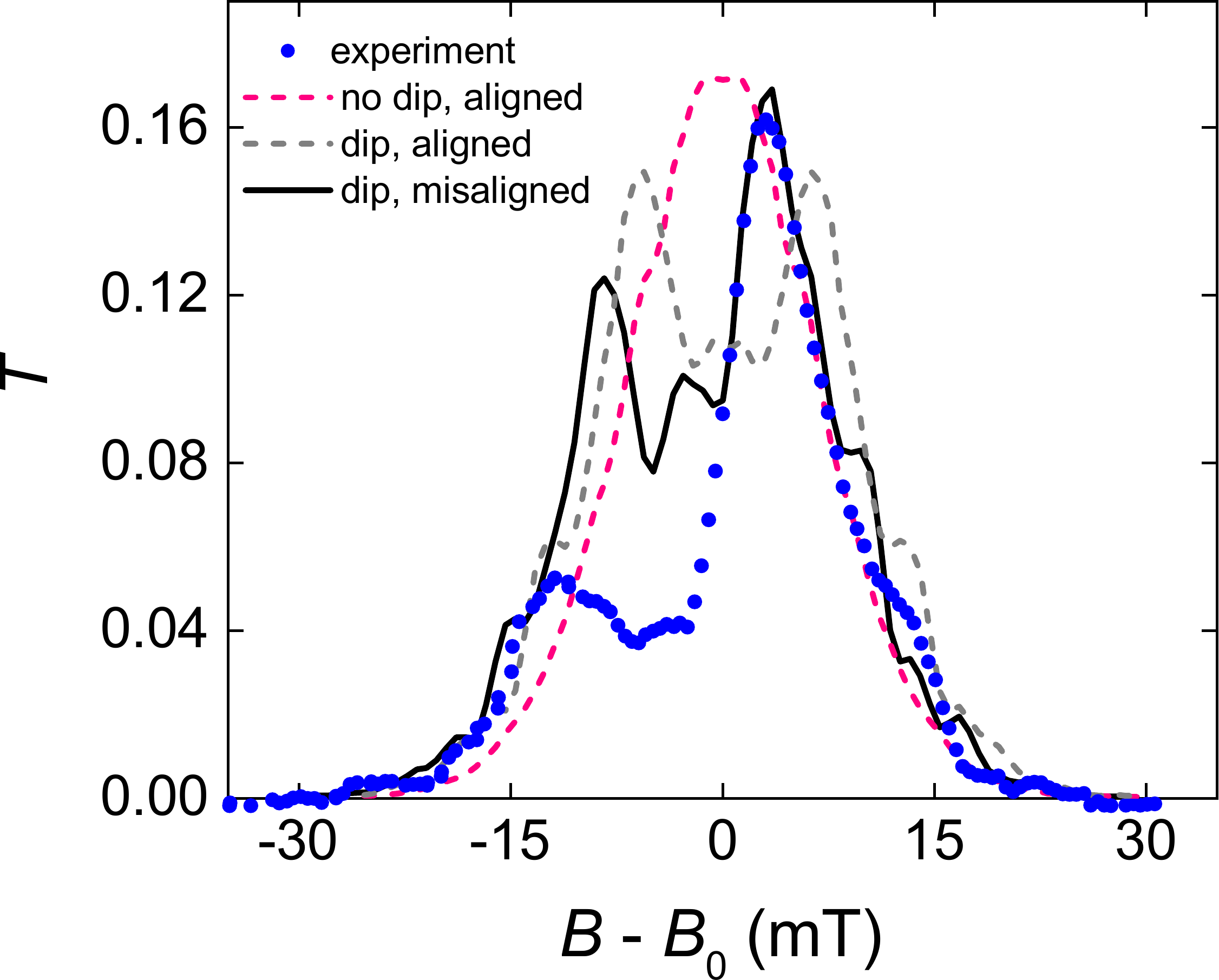}
\caption{Blue dots: measured magnetic-field dependent transmission through both QPCs at $N=1,M=7$, $T_{N=1,M=7}(B)$. Lines are calculated $T_{N=1,M=7}(B)$. Dashed magenta line: for a flat 2DES (without dip) and perfect alignment of the QPCs. Dashed gray line: including the potential dip for perfectly aligned QPCs. Black solid line: including both the potential dip and the misalignment of the QPCs with respect to the lens.
} 
\label{fig:Imp-vs-Imp+Dip}
\end{figure}}
we visualize the effect of the lateral misalignment of the QPCs with the lens as well as the potential dip in the center of the lens. The figure plots magnetic deflection data measured at $N=1$ and $M=7$ in comparison to our model prediction for the case of (i) perfect symmetry and no potential dip (pink dashed), (ii) perfect symmetry but the potential dip included (gray dashed), and (iii) QPCs simulated at their actual laterally shifted positions and potential dip included (black solid line). The dip splits the current peak centered at $B-B_0=0$ in two while the symmetry is preserved. The misalignment of the QPCs breaks the symmetry, in addition. The remaining deviations between model and measurement can be interpreted as our incomplete knowledge of the detailed electrostatic potential landscape. 

\section{Circuit characterization -- diffusive transport contributions}
\label{sec:circuit-charaterization}

The focus of our main article is on ballistic transport phenomena in the coherent transport regime. However, the ohmic resistances of the electrical leads inevitably cause diffusive carrier transport as well. We therefore need a clean separation between the ballistic and diffusive contributions to the current $I=\ibal+\idif$. In this section we demonstrate how we achieve such a separation. 

\begin{figure}[h!]
\includegraphics[width=0.8\columnwidth]{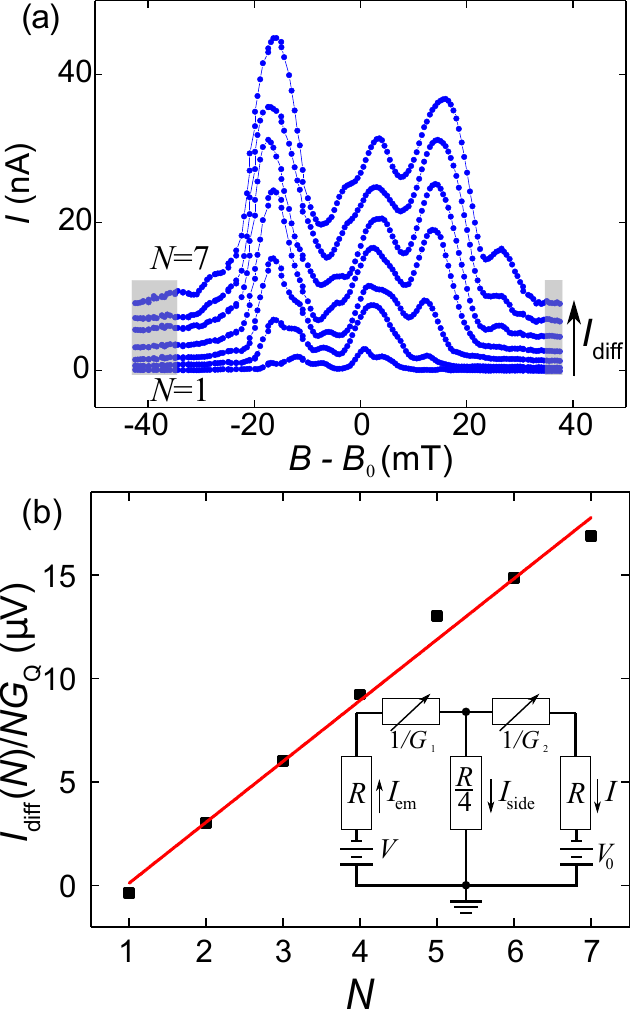}
\caption{(a) Measured detector current $I_{N,N}(B)$ flowing through QPC$_2$ for the source-drain voltage of $V=-1\,$mV  applied to the emitter, QPC$_1$. The region in between the QPCs is connected to ground. No lens is defined, $\vl=0$. The data correspond to one data set shown in Figure 2(a) of the main article but without artificial offsets. Both QPCs are tuned to the $N$th conductance plateau, respectively. (b) Inset: Simplified circuit diagram assuming diffusive transport. $R$ are the almost identical ohmic resistances of the leads, $G_{1,2}$ the tunable conductances of the QPCs’, $V$ the applied source-drain voltage and $V_0$ the input offset voltage of the current amplifier. Emitter, side and detector currents are labeled as $I_\text{em}$, $I_\text{side}$ and $I$ [corresponding to $I_{N,N}$ in (a)]. Black squares: Diffusive contribution of the detector current $\idif/NG_Q(N)$. \idif\ is the average value of $I_{N,N}$ within the magnetic field regions shaded in gray in panel (a). The red line is a linear least squares fit to the data. Using \eq{eq:Ioffapprox} it allows to extract the lead resistance $R=150\pm 7\,\Omega$ as well as the amplifier offset voltage $V_0=-2.8\pm0.6\,\mu$V. 
}
\label{fig:side_resistance-sketch}
\end{figure}

Both, $\ibal$ and $\idif$ can be  directly determined from our magnetic deflection measurements as displayed in Fig.\ 2(a) of the main article. In \fig{fig:side_resistance-sketch}a
we plot one of the sets of these $I(B)$ measurements for $N=M$. However, in contrast to the main article we plot the raw data without artificial offsets between the curves. There is no lens defined between the QPCs as $\vl=0$. At high enough magnetic fields, here for $|B-B_0|>35$\,mT, the cyclotron diameter of the ballistic electron orbits becomes smaller than the distance between the two QPCs, $2R_\text c=m^\star\vf/|e(B-B_0)|< l/2=2.3\,\mu$m, such that ballistic carriers emitted from one QPC can no longer reach the second QPC. This is the case in the regions shaded in gray in \fig{fig:side_resistance-sketch}a, hence, here the current still flowing through the detector QPC is diffusive (and independent of $B$). The diffusive current \idif\ is caused by the voltage drop across the detector, QPC$_2$. In the inset of \fig{fig:side_resistance-sketch}{b} we sketch a simplified circuit diagram of our setup which contains its ohmic resistances relevant in the diffusive transport regime. For simplicity we assume identical resistances, $R$, of all leads indicated in Fig.\ 1(a) of the main article as crossed squares. Note that independent characterization measurements of the sample with all gates grounded, i.e.\ no QPCs and no lens defined, yielded indeed approximately identical lead resistances of $R=144.2\pm 0.5\,\Omega$ for the six leads of the sample. These resistances are dominated by that of the ohmic contacts. $1/G_1$ and $1/G_2$ are the tunable resistances of the two QPCs, $V_0$ is the input offset voltage of the current amplifier and $V$ the applied source-drain voltage. Using Kirchhoff's circuit laws with the QPC conductances $G_1=G_2=NG_Q$ we find 
\begin{equation}\label{eq:Ioff}
\dfrac{\idif}{NG_Q}=\dfrac{XV-\left(5X+4\right)V_0}{6X^2+10X+4}\,;\quad X\equiv NG_Q R
\end{equation}
Since we are interested in $N\le7$ we find $NG_Q R<0.08\ll1$ and can approximate \eq{eq:Ioff} in first order with
\begin{equation}\label{eq:Ioffapprox}
\dfrac{\idif}{NG_Q}\simeq -V_0+\dfrac{V+5V_0}{4}NG_Q R
\end{equation}
In \fig{fig:side_resistance-sketch}{b} we present $\idif/(NG_Q)$ as a function of the mode number $N$ of the two QPCs using $V=-1$\,mV. The solid line is a linear fit which confirms \eq{eq:Ioffapprox}. The fit parameters are  $V_0=(-2.8\pm 0.6)\,\mu$V and $R=(150\pm 7)\,\Omega$, in good agreement with our preliminary characterization measurements.

To determine the conductance of the QPCs plotted in the pinch-off curves in Fig.\ 1(b) of the main article we likewise take into account the lead resistance $R$ by using
\begin{equation}
G=\left(\dfrac VI-2R\right)^{-1}\,,
\end{equation}
where $V$ is the voltage applied across the QPC in a two-terminal measurement and $I$ the corresponding measured current flowing through the QPC and its leads in series. The actual value of the lead resistance is thereby taken as a fit-parameter for each individual setup (depending on which leads are left floating and which QPC is being measured) to account for small variations in the resistances of the various leads.

We close this section by recalling our definition of the transmission through both QPCs in series, $T_{N,M}=I_{N,M}^\text{ball}/(G_\text Q V)$ with $G_\text Q=2e^2/h$. For $N=M$ we  write $T_{N,N}=I_{N,N}^\text{ball}/(G_\text Q V)$. Note, that we thereby subtract the diffusive current, $\ibal=I-\idif$, such that we compare only the contribution of electrons moving ballistically between emitter and detector. The transmission defined in terms of the ballistic current can therefore directly be compared with our model calculations.

\section{Further measurements}

\subsection{Magnetic deflection measurements}

\begin{figure}[h!]
\includegraphics[width=0.9\columnwidth]{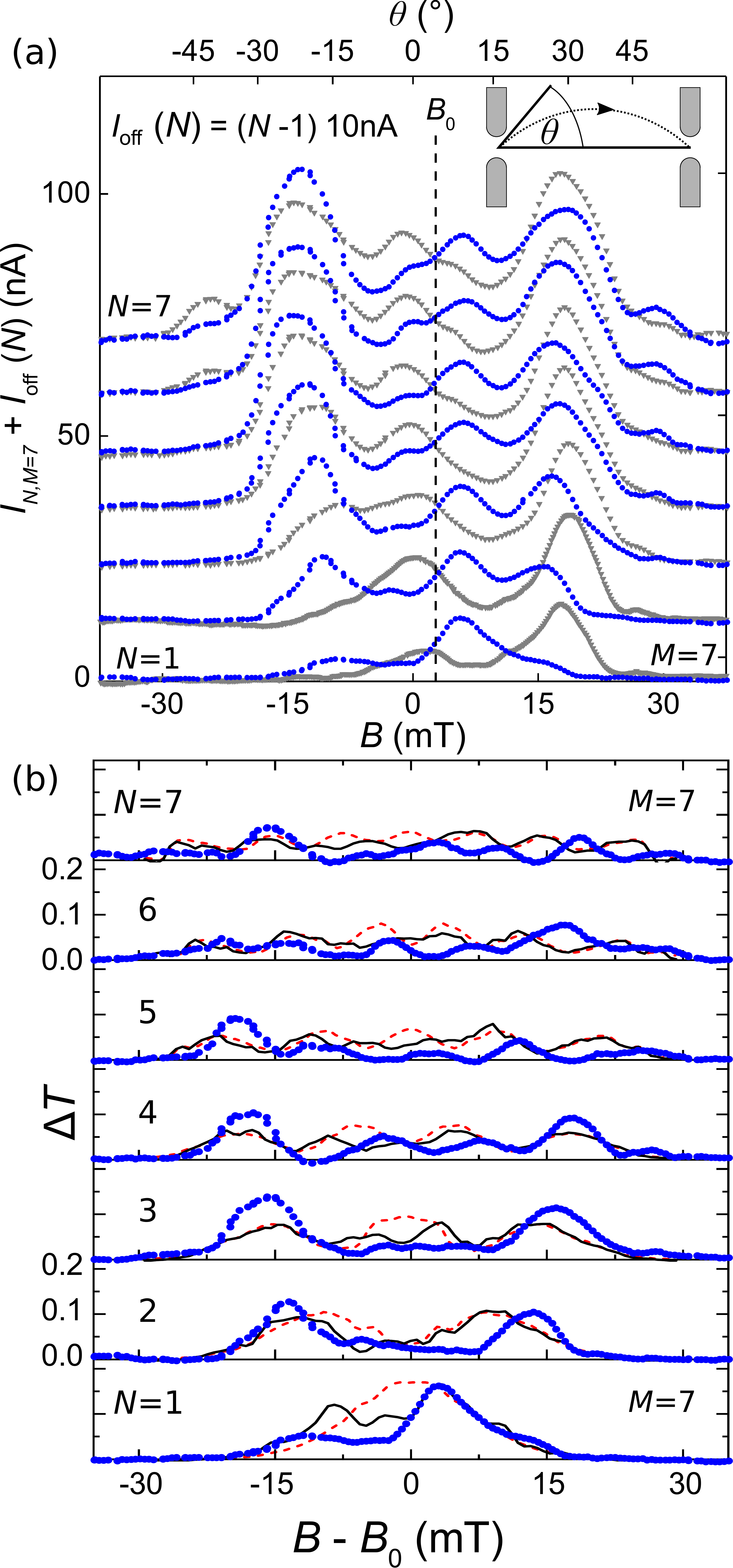}
\caption{
(a) Current through two QPCs in series with the detector fixed at $M=7$ while the emitter is opened from $N=1$ to $N=7$ as a function of the perpendicular magnetic field, $I_{N,M=7}(B)$. Data are vertically shifted for clarity. (b) Measured transmission differences $\Delta T_{N,M=7}=\left(I_{N,M=7}^\text{ball}-I_{N-1,M=7}^\text{ball}\right)/\left(G_\text QV\right)$ (blue dots); model curves $\Delta T_{N,M=7}=\sum_{m=1}^7 t_{N,m}$ (dashed lines) and calculated $\Delta T_{N,M=7}$ taking into account the potential dip at the lens and the misalignment of the QPCs and lens (solid black lines).}
\label{fig:T_N_M=7}
\end{figure}

For completeness we plot in \fig{fig:T_N_M=7}{a}
the raw data used for the transmission differences presented in Fig.\ 2(c) of the main article. The detector QPC is fixed at $M=7$ while we vary the emitter conductance between plateaus $1\le N\le7$.
In \fig{fig:T_N_M=7}{b} we plot the corresponding measured transmission differences $\Delta T_{N,M=7}=\left(I_{N,M=7}^\text{ball}-I_{N-1,M=7}^\text{ball}\right)/\left(G_\text QV\right)$ (blue dots, gray dots correspond to reversed current direction, cf. main article). Additionally, we show the corresponding calculated transmission differences $\Delta T_{N,M=7}=\sum_{m=1}^7 t_{N,m}$ with the pairwise transmission coefficients $t_{n,m}(B)$ as defined in the main article and $n=N$. We present the model curves for assuming a perfect (flat) 2DES (dashed magenta lines) and for the case of including the piezoelectric potential dip of the lens as well as the lateral misalignment of the QPCs (solid black lines) as discussed in Sections \ref{sec:self-consistent_calibration} and \ref{sec:nanostructure-shifts}.

%In \fig{fig:T_N_M=7}{c} we re-plot the raw data for $N=M$, namely $I_{N,N}(B)$, which are identical to those in Fig.\ 2(a) of the main article. By mutual subtraction from these data we determine the transmission differences $\Delta T_{N,N}=\left(I_{N,N}^\text{ball}-I_{N-1,N-1}^\text{ball}\right)/\left(G_\text QV\right)$ and plot them (blue dots) in panel (d). For completeness we also show the corresponding model curves, encoded as above. 
%
\begin{figure}[htb]
\includegraphics[width=0.8\columnwidth]{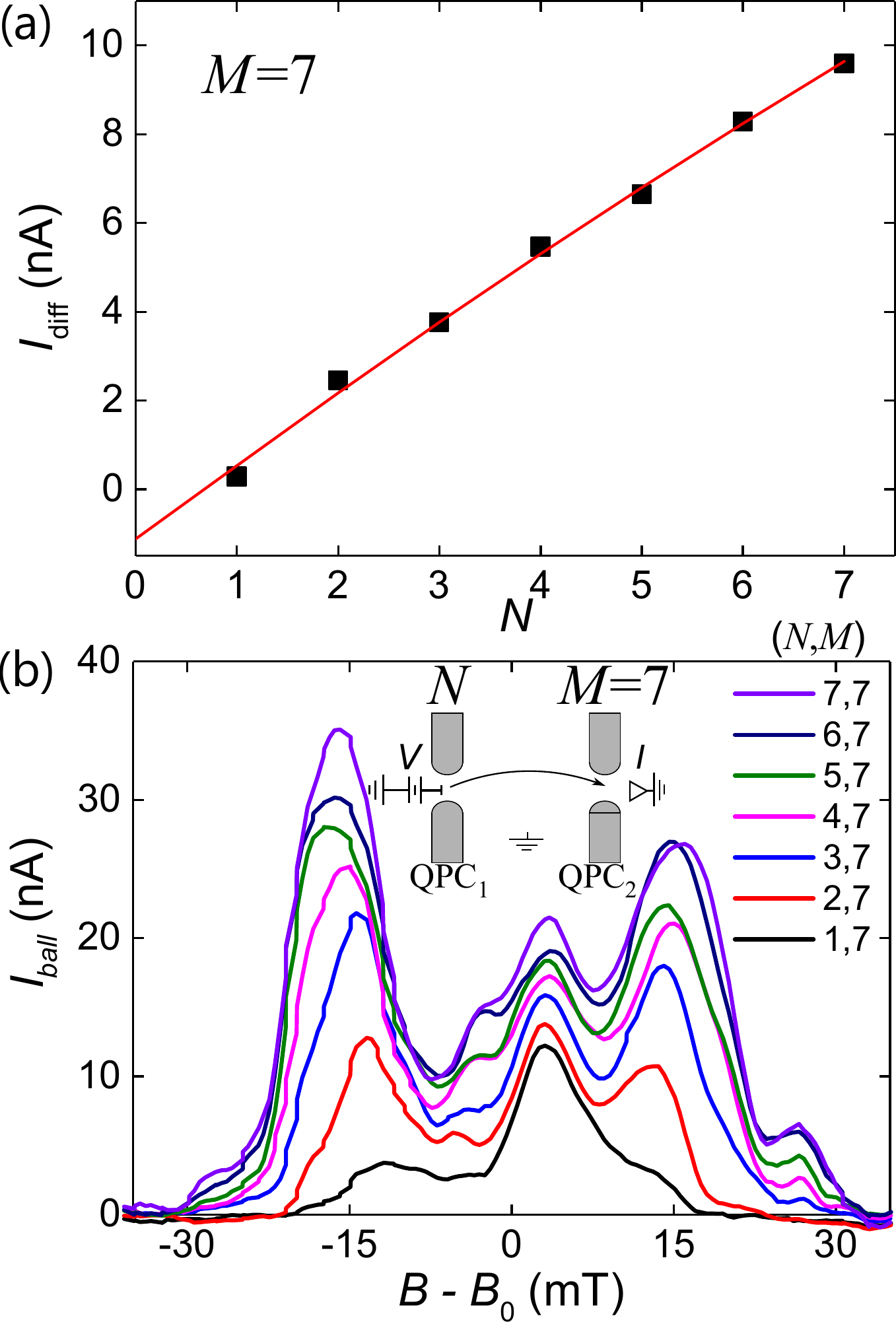}
\caption{Measured detector current $I(B)=\idif+\ibal(B)$ with the detector tuned to $M=7$ while the emitter is opened from $N=1$ to $N=7$.
(a) Diffusive current contribution $\idif$ (symbols). The line is a model curve according to \eq{eq:diffusive_current} using the lead resistance $R=150\,\Omega$, determined in  Sec.\ \ref{sec:circuit-charaterization} and the fit parameter $V_0=-1.07\,\mu$V. 
(b) Ballistic current $\ibal(B)$. The inset sketches the three-terminal measurement circuit. 
\label{fig:I(B)_det-M=7_em-N=1,...,7}}
\end{figure}
\begin{figure}[htb]
\includegraphics[width=0.85\columnwidth]{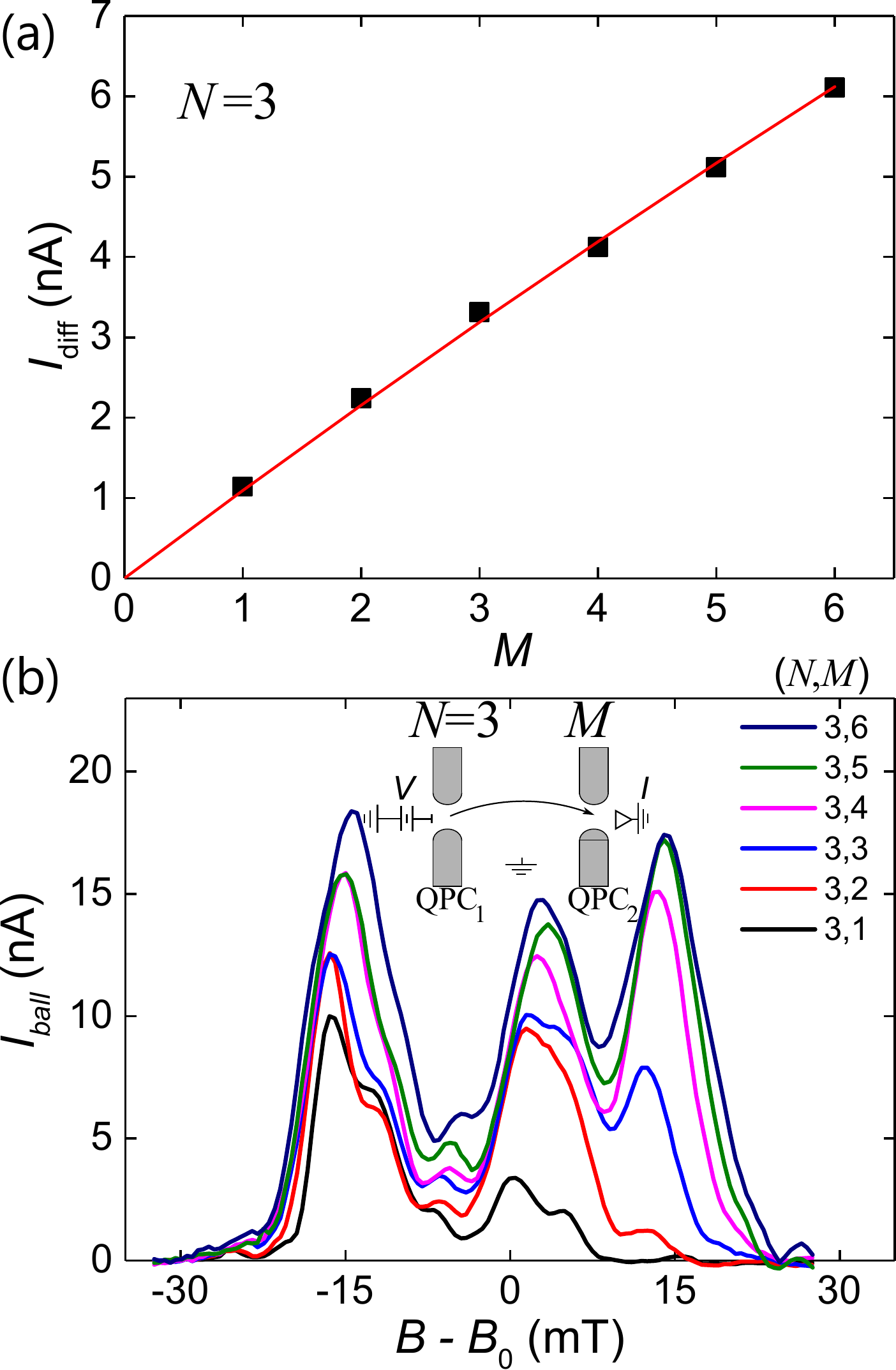}
\caption{Measured detector current $I(B)=\idif+\ibal(B)$ with the detector (QPC$_2$) opened from $M=1$ to $M=7$ while the emitter (QPC$_1$) remains at $N=3$.
(a) Diffusive current contribution $\idif$ (symbols). The line is a model curve according to \eq{eq:diffusive_current} using the lead resistance $R=150\,\Omega$, determined in  Sec.\ \ref{sec:circuit-charaterization} and the fit parameter $V_0=-5.96\,\mu$V. 
(b) Ballistic current $\ibal(B)$. The inset sketches the three-terminal measurement circuit.
\label{fig:I(B)_em-N=3_det-M=1,...,7}}
\end{figure}
In \fig{fig:I(B)_det-M=7_em-N=1,...,7}{b} we replot the blue data set shown \fig{fig:T_N_M=7}{} after subtracting the diffusive current contribution, following the procedure introduced in Sec.\ \ref{sec:circuit-charaterization}, which we plot in \fig{fig:I(B)_det-M=7_em-N=1,...,7}{a} as a function of $N=G_\text{emitter}/G_Q$. In this measurement $\vl=0$, while we apply the voltage $V=-1$\,mV across the emitter QPC with the intermediate region being grounded, see inset in panel (b). Generalizing \eq{eq:Ioff} for the case of $N\ne M$ and for $G_1=NG_Q$ being the emitter and $G_2=MG_Q$ the detector conductance we can predict the diffusive current contribution from Kirchhoff's law for the circuit sketched in \fig{fig:side_resistance-sketch}b to be
\begin{equation}\label{eq:diffusive_current}
\idif=G_Q M\dfrac{VRG_Q N-V_0\left(5RG_Q N+4\right)}{6(RG_Q)^2NM+5RG_Q(N+M)+4} 
\end{equation}
where $R=150\,\Omega$ was determined in Sec.\ \ref{sec:circuit-charaterization}, while the offset voltage of the current amplifier $V_0$ is again fit-parameter as it can slightly drift between measurements. The solid line in panel (a) is a model curve according to \eq{eq:diffusive_current} for $V_0=-1.07\mu$V. In panel (b) we present the ballistic current $\ibal(B)$. As expected $\ibal(B)$ increases with the emitter conductance, reflecting the mode structure of the system of coupled QPCs. 

In \fig{fig:I(B)_em-N=3_det-M=1,...,7}{} we plot comparable data, but for the emitter (QPC$_1$) tuned to $N=3$ and the detector with varying conductance $1\le M\le7$. The solid line in panel (a) is a model curve according to \eq{eq:diffusive_current} for $V_0=-5.96\mu$V. The apparent differences between $\ibal(B)$ for fixed $M=7$ versus fixed $N=3$ can be explained in terms of different elements of the transmission matrix contributing to the ballistic current, cf.\ Sec.\ \ref{sec:matrices}. This, however, is beyond the scope of the present article.

\subsection{Electrostatic focusing experiments}

In \fig{fig:Lens-M=7,N=1,...,7}{a}
\begin{figure}[t]
\includegraphics[width=0.85\columnwidth]{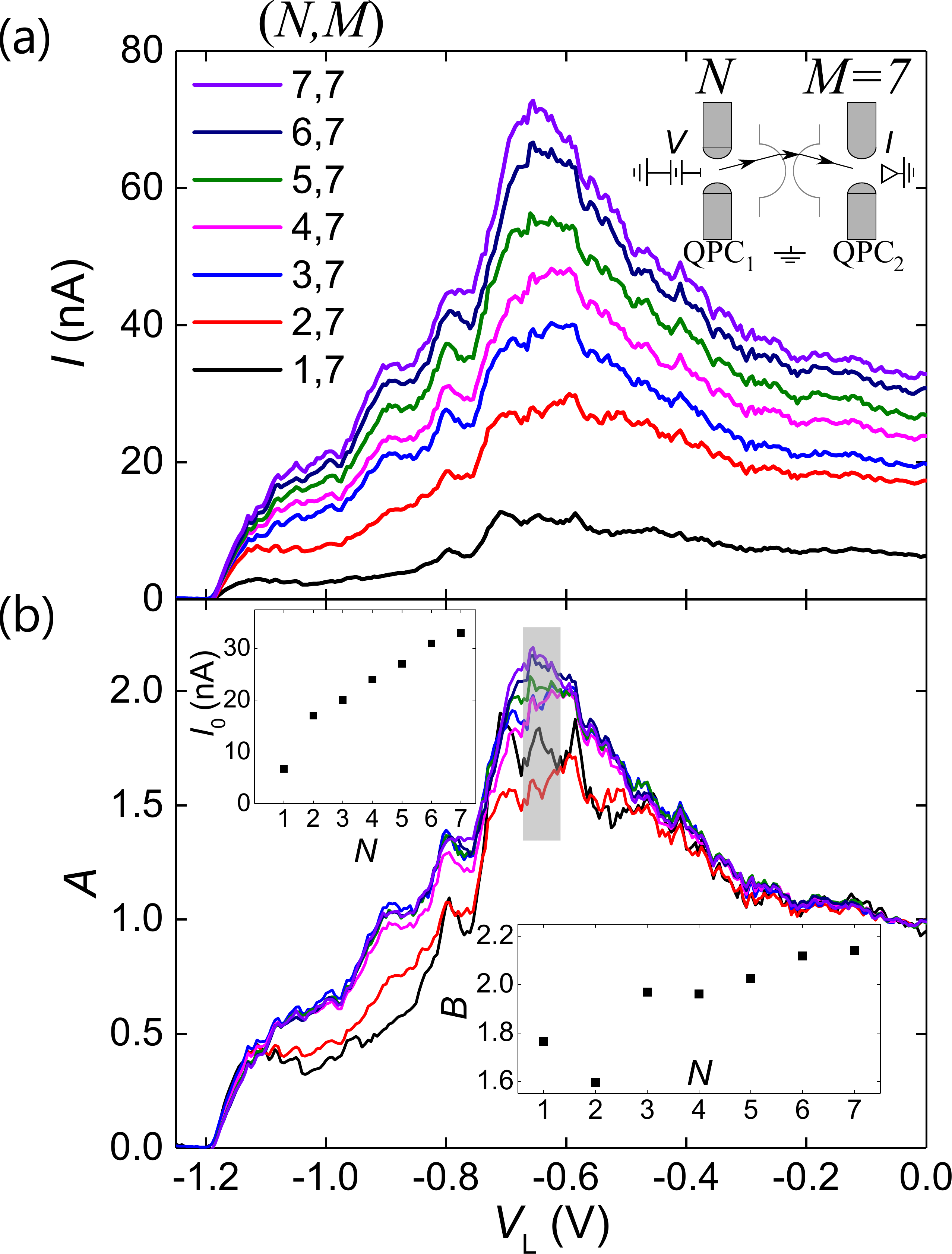}
\caption{(a) Measured detector current $I(\vl)$ at $B=B_0$ with the detector (QPC$_2$) tuned to $M=7$ while the emitter (QPC$_1$) is opened from $N=1$ to $N=7$ from the bottom to top. The inset sketches the three-terminal measurement circuit.
(b) Same data but normalized to $I_0=\left<I(\vl>-0.1)\right>$. Upper left inset: $I_0(N)$. Lower right inset: Maximum current enhancement compared to $I_0$ averaged over the shaded region -0.66\,V$<\vl<-0.62$\,V. 
\label{fig:Lens-M=7,N=1,...,7}}
\end{figure}
we use QPC$_1$ as emitter and vary its conductance through $N=1\dots7$ while we use QPC$_2$ tuned to $M=7$ as detector. Here we plot the detector current $I(\vl)$ at $B=B_0$ instead of sweeping $B$ as above. In panel (b) we present the same data but normalized to the current near $\vl=0$, i.e.\ $I(\vl)/I_0$ with $I_0=\left<I(\vl>-0.1)\right>$. For completeness we also plot $I_0(N)$ in the upper left inset. In the lower right inset of panel (b) we instead display the current maximum $I_\text{max}/I_0$ averaged over the shaded region of \vl\ as a function of $N$. Besides strong fluctuations for $N\le3$, the ratio $I_\text{max}/I_0$ increases with $N$. This behavior demonstrates that focusing works better for larger $N$. We expect this trend, because (i) the dip in the center of the lens  potential near the optical axis leads to additional scattering. For higher modes, the current density shifts more and more to the flanks of the Hermite-Gaussian beam such that most carriers avoid the dip and hit the outer parts of the lens.

In \fig{fig:I(B,VL)-em_N=2-det_M=7}{}
\begin{figure}[htb]
\includegraphics[width=1\columnwidth]{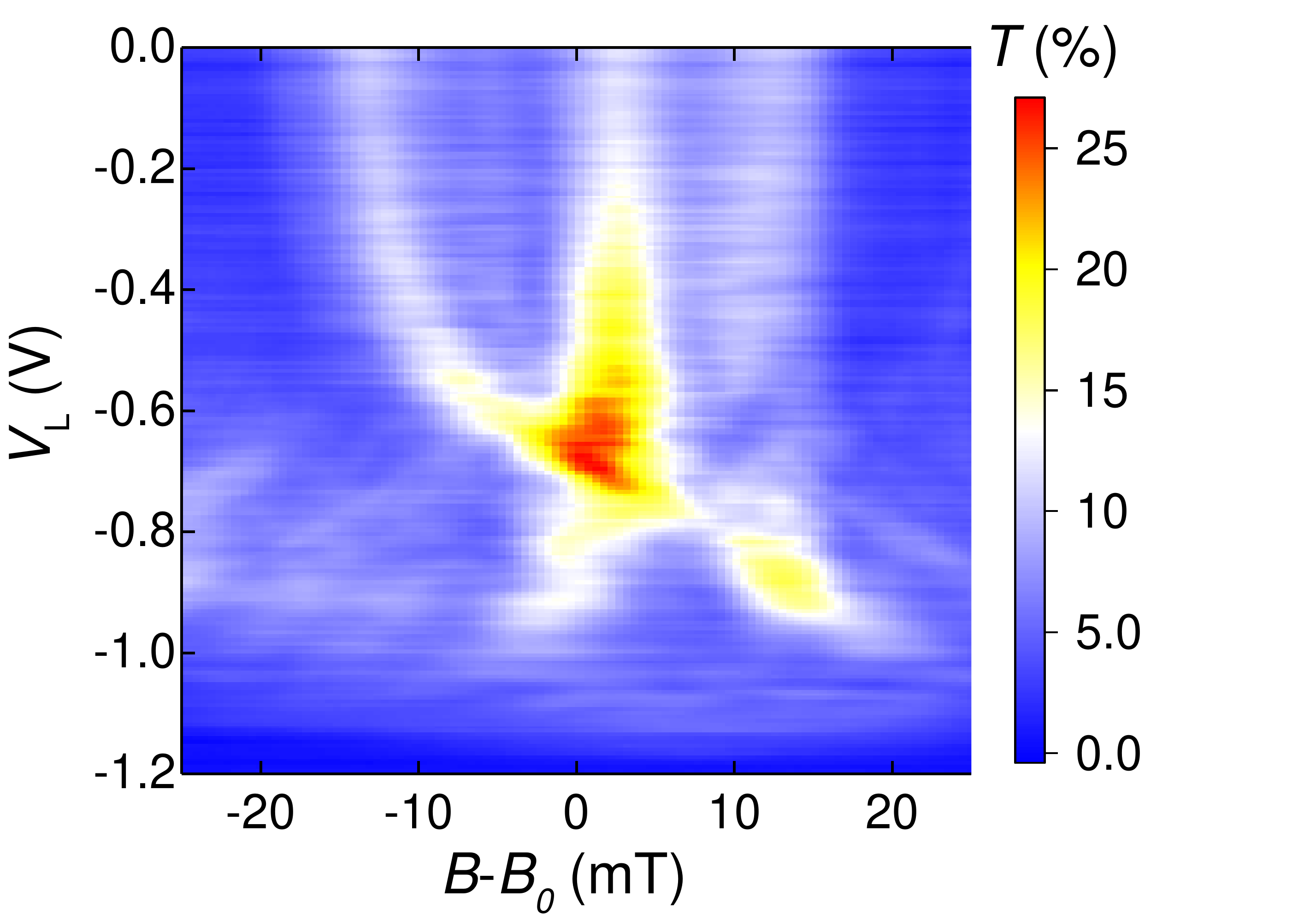}
\caption{Transmission $T(B,\vl)$ at $B=B_0$ with the detector (QPC$_2$) tuned to $M=7$ while the emitter (QPC$_1$) is tuned to $N=2$.
\label{fig:I(B,VL)-em_N=2-det_M=7}}
\end{figure}
we show a focusing experiment plotting $I(B,\vl)$ as in Fig.\ 3(a) of the main article but for $N=2$ and $M=7$ instead of $N=M=7$. The figure demonstrates that focusing works independently of the detailed mode structure of the QPCs. In particular, the focusing condition is still best fulfilled for $\vl=-0.64\,$V as the lens properties are independent of the current density profile.

\subsection{Possible influence of disorder between the two QPCs}
\label{sec:disorder-potential}

Disorder can influence the transport properties of mesoscopic devices even if the mean-free-path exceeds the device size by far. In this regime we expect small angle scattering, e.g.\ originated by charged defects such as the ionized donor atoms, to have a small influence on the quantum mechanical phase of the carrier dynamics. However, even a single hard-wall, i.e., large angle scatterer can alter the ballistic properties completely by reflecting carriers and thereby generating an uncontrolled  standing waves pattern. Above, we have demonstrated that our experimental results can be explained by accounting for the existing geometric imperfections, in our case a slight misalignment of the QPCs, and an imperfect lens caused by a potential dip related with the piezoelectric effect.

\begin{figure}[tbh]
\includegraphics[width=0.75\columnwidth]{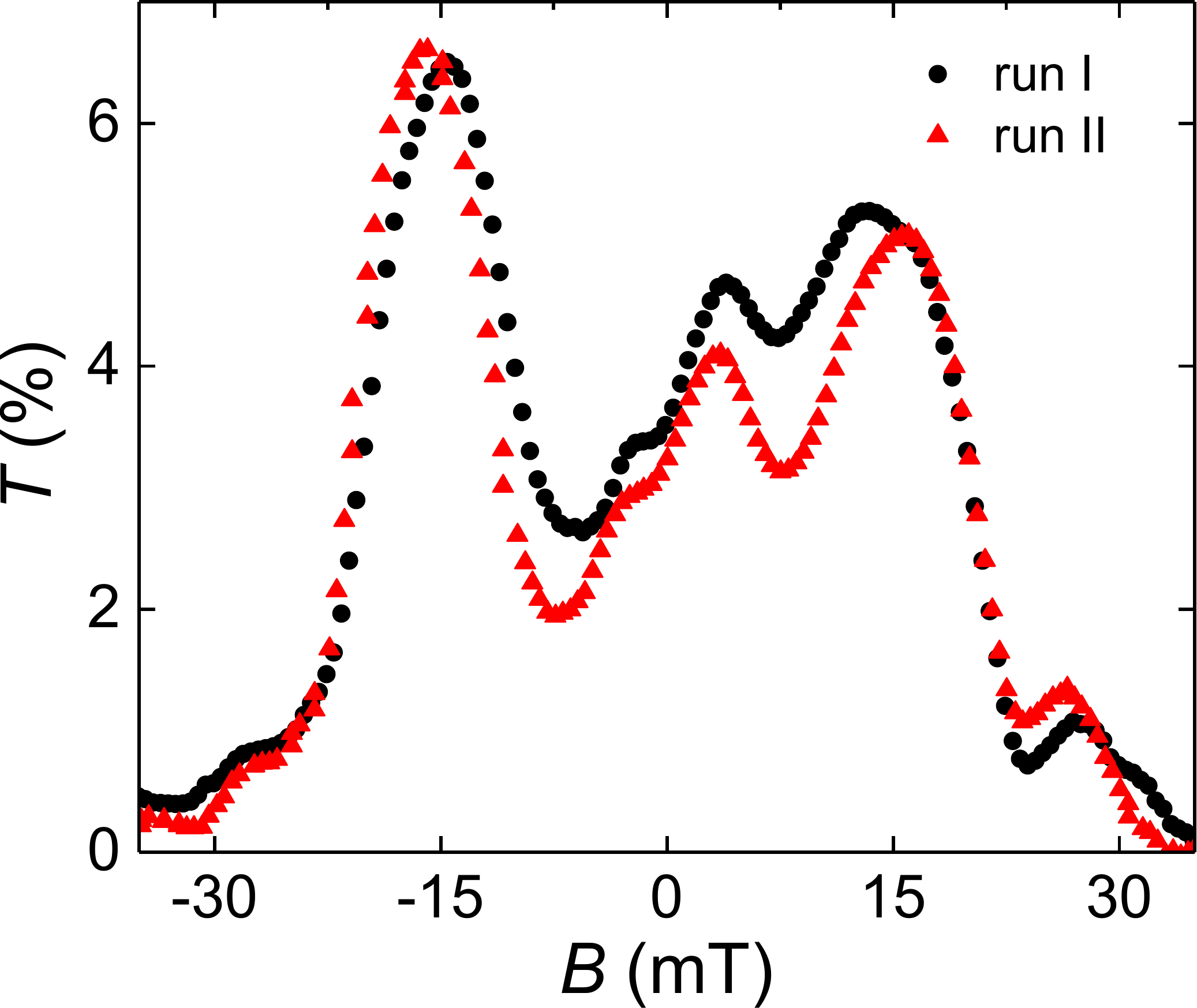}
\caption{Serial transmission $T(B)$ through both QPCs, tuned to $N=M=7$. Black dots: measured $T(B)$ in run I. Red triangles: same measurement after warming up, illuminating and cooling down the sample again (run II).}
\label{fig:modestructure-different_runs}
\end{figure}

To experimentally explore the influence of disorder, we performed the some of our measurements twice in separate runs. However, between the two runs we warmed the sample up to room temperature, illuminated it with daylight and than cooled it back down. Because this procedure affects many defects by excitation and diffusion, the potential landscape induced by defects should vary between the two runs.  We present a typical result in \fig{fig:modestructure-different_runs}{},
which shows the measured transmission through both QPCs as a function of the magnetic field for $N=M=7$. The two curves are very similar with identical features for both runs. We conclude, that the traces $T(B)$ are not strongly influenced by disorder, at most by energetically stable defects which remained unchanged despite our resetting procedure. We tested this scenario and found, that we could describe our measurements using an alternative scenario, assuming a perfect sample (without misalignments and with a flat lens potential) but which contains a single hard-wall scatterer. To completely exclude the possibility of disorder related influences we had to repeat our measurements but using a perfect sample.

\subsection{Effect of a global topgate}

\begin{figure}[htb]
\includegraphics[width=0.75\columnwidth]{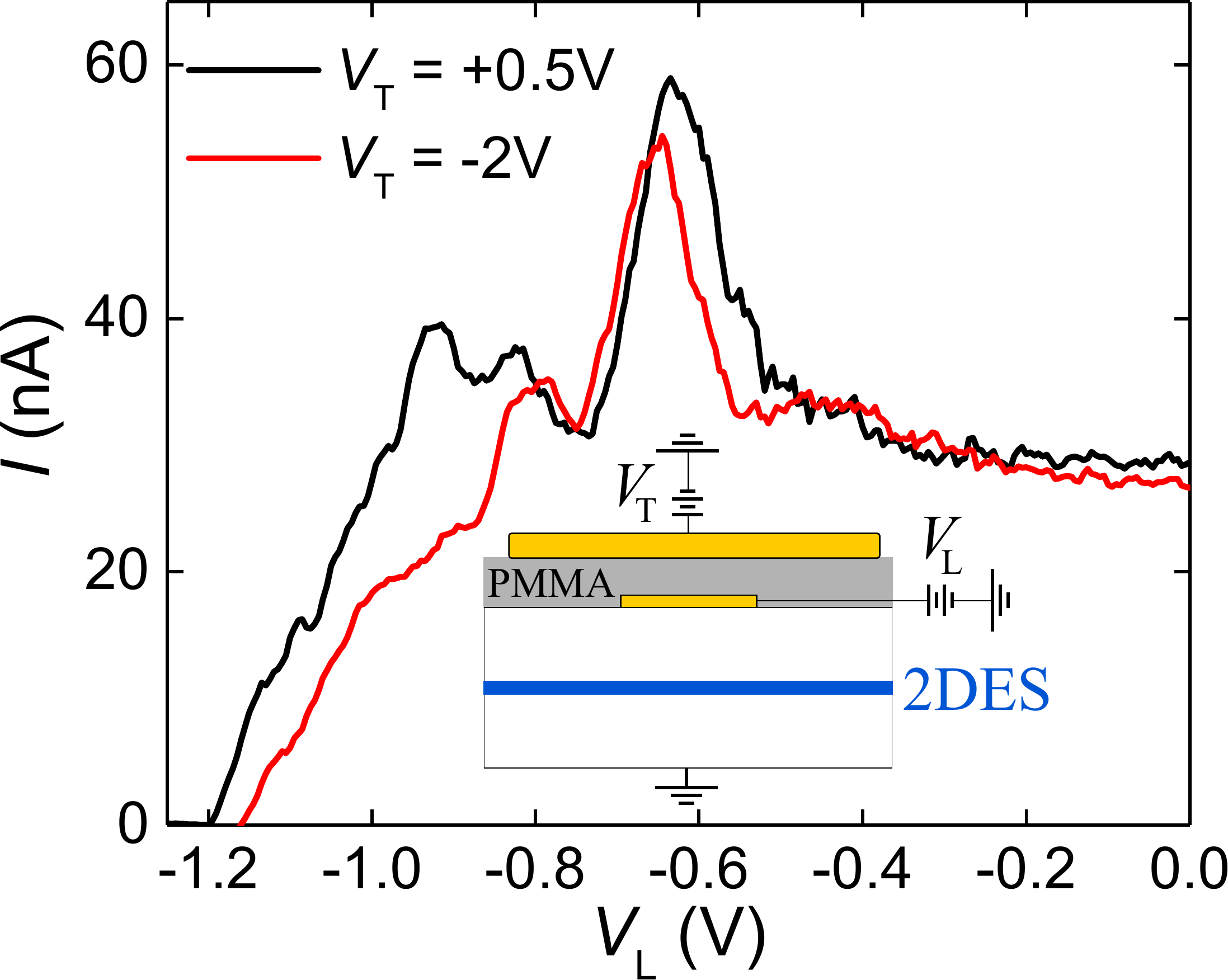}
\caption{Measurements with a voltage $V_\text{T}$ applied to a global topgate which covers the entire structure of interest, i.e.\ the QPCs and the lens. The inset sketches the setup. The main panel shows the detector (QPC$_1$) current $I(\vl)$ at $B=B_0$ for $N=M=7$ for $V_\text{T}=+0.5$\,V (black line) versus $V_\text{T}=-2$\,V (red).
\label{fig:topgate-measurement}}
\end{figure}

Finally, we study the effect of a global top gate, which covers the entire structure of interest including both QPCs and the lens gate. The top gate is electrically isolated and separate from the gates defining the QPCs and the lens by means of an approximately 130\,nm thick layer of cross-linked PMMA, see inset of \fig{fig:topgate-measurement}{}.
By applying a voltage $V_\text{T}$ to the topgate, we can vary the carrier density, hence, the Fermi energy $\ef^0$ of the 2DES. In \fig{fig:topgate-measurement}{} we plot two curves $I(\vl)$ measured at different values of $V_\text{T}$. They display two clear differences: (i) a more negative $V_\text{T}$ causes a decreased carrier density. This can be seen in terms of a shift of the pinch-off point, in our case by $\simeq40\,$mV. If the lens were not affected by the top-gate voltage, the focus point should shift in the same direction and by the same amount along \vl. In contrast, (ii) the focus point is shifted in the opposite direction than the pinch-off point. It indicates, that in the plane of the 2DES the electric field distribution beneath the lens gate is influenced not only by \vl\ but also by $V_\text{T}$ (in particular along the edge of the lens gate). Because this field distribution defines the lens properties, it demonstrates that a global top gate can be used to fine tune the properties of an electrostatic lens.

%%%%%%%%%%%%%%%%%%%%%%%%%%%%%%%%%%%%%%%%%%%%%%%%%%%%%%%%%%%%%%%%%%%%%
\bibliography{literature,../../../zitate/zitate}

\end{document}